\documentclass[12pt]{article}
\pdfoutput=1

\usepackage{authblk} 
\usepackage{geometry} 
\newgeometry{vmargin={25mm}, hmargin={28mm,28mm}}

\usepackage{amsthm, amsfonts}
\usepackage{mathrsfs, mathtools}
\usepackage{physics}

\usepackage{subcaption} 
\usepackage{graphicx}

\usepackage{tikz}
\usepackage{tikz-cd}
\usepackage{pgfplots}
\pgfplotsset{compat=1.14}

\usepackage[hidelinks]{hyperref}
\hypersetup{
  colorlinks   = true, 
  linkcolor={red!50!black},
  citecolor={blue!50!black},
  urlcolor={blue!80!black}
}
\usepackage{cleveref}
\usepackage{natbib}

\usepackage{multirow} 
\usepackage{relsize} 
\usepackage{siunitx} 
\usepackage{float}

\title{\LARGE{\textbf{Yet anOther Dose Algorithm (YODA) for independent computations of dose and dose changes due to anatomical changes}}}

\author[1,2]{Tiberiu Burlacu}
\author[1,2]{Danny Lathouwers\footnote{Both authors contributed equally}}
\author[1,2]{Zolt\'{a}n Perk\'{o}\protect\footnotemark[1]}

\affil[1]{\footnotesize{Delft University of Technology, Faculty of Applied Sciences, Delft, The Netherlands}}
\affil[2]{\footnotesize{HollandPTC consortium\footnote{HollandPTC consortium – Erasmus Medical Center, Rotterdam, Holland Proton Therapy Centre, Delft, Leiden University Medical Center (LUMC), Leiden and Delft University of Technology, Delft, The Netherlands}, Delft, The Netherlands}}

\date{\small{\today}}

\begin{document}

\maketitle

\begin{abstract}
  \textbf{Purpose:} To assess the viability of a physics-based, deterministic and adjoint-capable algorithm for performing treatment planning system independent dose calculations and for computing dosimetric differences caused by anatomical changes.

  \textbf{Methods:} A semi-numerical approach is employed to solve two partial differential equations for the proton phase-space density which determines the deposited dose. Lateral hetereogeneities are accounted for by an optimized (Gaussian) beam splitting scheme. Adjoint theory is applied to approximate the change in the deposited dose caused by a new underlying patient anatomy.

  \textbf{Results:} The quality of the dose engine was benchmarked through three-dimensional gamma index comparisons against Monte Carlo simulations done in TOPAS. The worst passing rate for the gamma index with (\SI{1}{\milli \meter}, \SI{1}{\percent}, \SI{10}{\percent} dose cut-off) criteria is \SI{95.62}{\percent}. The effect of delivering treatment plans on repeat CTs was also tested. For a non-robustly optimized plan the adjoint component was accurate to \SI{6.2}{\percent} while for a robustly optimized plan it was accurate to \SI{1}{\percent}.

  \textbf{Conclusions:} YODA is capable of accurate dose computations in both single and multi spot irradiations when compared to TOPAS. Moreover, it is able to compute dosimetric differences due to anatomical changes with small to moderate errors thereby facilitating its use for patient-specific quality assurance in online adaptive proton therapy.

\end{abstract}

\vspace{2pc}
\maketitle

\section{Introduction}

\subsection{Online Adaptive Proton Therapy and Quality Assurance}

Proton Therapy (PT) promises to improve on conventional photon based radiotherapy for curative cancer treatments due to the characteristics of its dose-depth curve. The proton dose-depth curve shows simultaneously lower doses achievable in organs at risk (OARs) and an increased target dose conformality due to the presence of the Bragg peak (BP) \citep{paganettiProtonTherapyPhysics2016}. Both target dose conformality and dose in OARs can however be degraded by the presence of uncertainties. Typical examples of uncertainty sources are the conversion of Hounsfield units (HU) in Computed Tomography (CT) scans to proton stopping powers, the daily positioning of the patient in the treatment room or the short and long-term anatomical changes occuring in the patient \citep{lomaxIntensityModulatedProton2008,lomaxIntensityModulatedProton2008a}. To improve target coverage, clinical proton plans are subjected to robust optimization \citep{voortRobustnessRecipesMinimax2016}. Robust optimization seeks to create plans that perform well under a number of error scenarios such as range and patient set-up errors \citep{unkelbachRobustProtonTreatment2018}. In doing so, robust optimization creates a high dose margin around the target in the surrounding OARs \citep{vandewaterPriceRobustnessImpact2016}.
While this makes treatments less sensitive to the included range and setup errors, other scenarios (e.g., weight loss over the course of week long treatments) are too complex to be modelled in a straightforward manner \citep{paganettiAdaptiveProtonTherapy2021}.

The workflow of Online Adaptive Proton Therapy (OAPT) would allow the reduction of the complexity and number of robust optimization scenarios. In this workflow, a new daily CT scan of the patient is acquired, a new fully re-optimized treatment plan is quickly created and thereafter safely delivered \citep{botasOnlineAdaptionApproaches2018}. This workflow would avoid tumor underdosage and would result in a lowering of the Normal Tissue Complication Probability (NTCP) through the reduction of the necessary margins around the tumor down to the intra-fractional ones for a robustly optimized plan \citep{paganettiAdaptiveProtonTherapy2021}. Unfortunately, the computational expense of plan re-optimization \citep{menGPUbasedUltrafastDirect2010} and the time needed for the (mostly manual) plan quality assurance (QA) process \citep{barrettPracticalRadiotherapyPlanning2009} have so far rendered this workflow practically infeasible.

One bottleneck in both plan re-optimization and the QA process is the lack of accurate and fast treatment plan dose computations. In intensity modulated proton therapy (IMPT) plans typically hundreds or even thousands of pencil beams (PBs) or spots have their fluence (and thereby dose) modulated during the optimization process \citep{schwarzTreatmentPlanningProton2011}. Machine QA (ranging from daily to yearly) procedures entail a series of time-consuming measurements meant to assess the constancy of beam properties and the correct functioning of its delivery system \citep{arjomandyOverviewComprehensiveProton2009,liUseTreatmentLog2013}. In addition patient-specific quality assurance (PSQA) must also be performed, with the goal to assess whether the differences between the planned and delivered dose distributions are within the clinically acceptable range of \SI{\pm 3}{\percent}\citep{gottschalkPassiveBeamSpreading2004} and to perform an independent check of the patient-specific dose that the treatment planning system (TPS) computes \citep{johnsonHighlyEfficientSensitive2019}. Additionally, PSQA also functions as a redundant check of the machine function \citep{frankProtonTherapyIndications2020}. Currently, PSQA is manually performed via dosimetric measurements which are infeasible in an OAPT workflow. TPS independent dose calculations (IDCs) based on log-files (records of the delivered spot positions and corresponding Monitor Units (MUs)) have been proposed as a solution for automating PSQA \citep{liUseTreatmentLog2013}. They have been shown to have similar accuracy to dosimetric measurements and could yield clinically relevant metrics \citep{meierIndependentDoseCalculations2015, meijersFeasibilityPatientSpecific2020}. Such an approach has potential within the time-constrained workflow of OAPT and could also increase clinical throughput \citep{meijersFeasibilityPatientSpecific2020} by reducing the time spent on QA.

\subsection{A hybrid independent dose computation approach}
\label{subsec:yoda_intro}

To perform fast, TPS independent and log-file based dose computations the interactions between the proton beam and the patient must be modelled, ideally not only using a different implementation of the TPS dose engine but also using a different methodology altogether. The two methods that are likely to be employed by a TPS are the Monte Carlo (MC) method and the analytical PB method. The MC method (e.g., TOPAS \citep{perlTOPASInnovativeProton2012}) trades fast computation times for high computational precision \citep{zheng-mingOverviewTransportTheory1993} by solving the in-tissue proton balance equation (i.e. the Linear Boltzmann Equation) using statistical sampling methods. The analytical PB method (e.g., Bortfeld's model \citep{bortfeldAnalyticalApproximationBragg1997}) trades high precision for fast computation times by employing a series of approximations and fits to obtain the dose in the tissue of interest. PB methods are still routinely used in TPS \citep{trnkovaFactorsInfluencingPerformance2016} despite their limitations being well documented \citep{soukupPencilBeamAlgorithm2005}.

We previously presented a methodologically different approach based on a deterministic solution of the Linear Boltzmann Equation \citep{burlacuDeterministicAdjointBasedSemiAnalytical2023a}. This approach, which will henceforth be referred to as Yet anOther Dose Algorithm (YODA), is a hybrid numerical and analytical solution to a physics motivated approximation of the same equation that MC methods solve. The method strikes a balance in terms of accuracy versus speed. It is accurate with respect to MC methods due to the physical modelling of the interactions between the proton beam and the patient and it is fast due to the partly analytical solution. An additional advantage of this approach is the ease of applying the adjoint method. Given planning and repeat CT images with delineated structures and a treatment plan the adjoint method computes an approximation of the change in dose caused by delivering the treatment plan to the repeat CT image, thereby avoiding an expensive re-computation. 

The purpose of this work is to demonstrate and test YODA's performance in real anatomies. Thus, YODA is compared to TOPAS in several irradiation sites and the adjoint engine's capability of accurately computing dose changes due to anatomy changes is benchmarked using TPS generated irradiation plans. This paper also documents the improvements brought to YODA, i.e., a more stable and an order more accurate numerical integration method, a better elastic scattering model for the proton beam, improved modelling in the Fermi-Eyges equation, a laterally optimized Gaussian beam splitting scheme and RT DICOM clinical treatment plan reading, compared to the original documented version \citep{burlacuDeterministicAdjointBasedSemiAnalytical2023a}. The details of these changes next to the theoretical framework of YODA are given in Section \ref{sec:methods}. Section \ref{sec:results_and_discussion} presents the results and their discussion while Section \ref{sec:conclusion} presents the conclusions and future outlook.

\section{Methods}
\label{sec:methods}

\subsection{Algorithm components}

To model the proton phase-space density in the patient the integro-differential Linear Boltzmann Equation which all MC methods are based on is simplified using physics based approximations. The approximations employed, namely the continuous slowing down approximation, the energy-loss straggling approximation, the small-angle Fokker-Planck approximation \citep{burlacuDeterministicAdjointBasedSemiAnalytical2023a}, result in two partial differential equations (PDEs) that describe the proton phase-space density in an in-depth hetereogeneous and laterally homogeneous geometry. The first PDE is the one-dimensional Fokker-Planck (FP) equation,
\begin{align}
    \label{eq:1DFP}
    \text{1DFP}(\varphi_{FP}) = \pdv{\varphi_{FP}}{z}
    - \pdv{S(z, E) \varphi_{FP}}{E}
    - \frac{1}{2} \pdv[2]{T(z, E) \varphi_{FP}}{E}
    + \Sigma_a(z, E) \varphi_{FP} = 0,
\end{align}
with $\varphi_{FP} = \varphi_{FP}(z, E)$ the proton Fokker-Planck flux that depends on the depth along the central axis of the beam $z \in \mathbb{R}$ and on the beam energy $E \in \mathbb{R}$, $S(z, E)$ the proton stopping power, $T(z, E)$ the energy straggling coefficient and $\Sigma_a(z, E)$ the macroscopic absorption cross section. The second equation is the Fermi-Eyges equation,
\begin{align}
    \label{eq:FE}
    \Upsilon\qty(\varphi_{FE}) = \pdv{\varphi_{FE}}{z}
    + \Omega_x \pdv{\varphi_{FE}}{x} + \Omega_y \pdv{\varphi_{FE}}{y}
    - \overline{\Sigma_{tr}}(z) \qty(\pdv[2]{\varphi_{FE}}{\Omega_x}
    + \pdv[2]{\varphi_{FE}}{\Omega_y}) = 0,
\end{align}
with $\varphi_{FE}=\varphi_{FE}(\vu*{\Omega}, \vb*{r})$ the Fermi-Eyges flux, $\vu*{\Omega} = (\Omega_x, \Omega_y) \in \mathbb{R}^2$ the direction cosines along the x and y proton velocity axes, $\vb*{r} = (x, y, z) \in \mathbb{R}^3$ a point in physical beam-eye view space (with z the depth along the beam) and $\overline{\Sigma_{tr}}(z)$ the energy spectrum (i.e., $\varphi_{FP}$) averaged macroscopic transport cross section. The individual solutions to equations \ref{eq:1DFP} and \ref{eq:FE} are multiplied to obtain the complete 6-dimensional proton phase-space density,
\begin{equation}
    \varphi(\vb*{r}, \vu{\Omega}, E) = \varphi_{FE}(\vb*{r},\vu*{\Omega}) \cdot \varphi_{FP}(z,E).
\end{equation}
All clinically relevant metrics, such as the dose distribution or the NTCP, can be derived from the proton phase-space density $\varphi$.

\subsubsection{The Fokker-Planck equation}

The one-dimensional Fokker-Planck equation \ref{eq:1DFP} is numerically solved using the Symmetric Interior Penalty Galerkin (SIPG) \citep{riviereDiscontinuousGalerkinMethods2008} method in the energy domain and the three-stage, third-order accurate Singly Diagonally Implicit Runge-Kutta (SDIRK) method \citep{kennedyDiagonallyImplicitRungeKutta2016} in space (depth). The energy domain is discretized into $N_g$ intervals called groups. Within each group $\varphi_{FP}$ is approximated as an expansion around the first three group-centered Legendre polynomial basis functions resulting in a method that is third order accurate in energy. The one-dimensional Fokker-Planck equation \ref{eq:1DFP} is supplemented with boundary conditions in energy (BCE) and space (BCS),
\begin{align}
    \text{BCE: } & \eval{\varphi_{FP}(z, E)}_{E=E_{max}} = 0,
    \eval{\pdv{\varphi_{FP}(z, E)}{E}}_{E=E_{max}} = 0,
    \eval{\pdv{\varphi_{FP}(z, E)}{E}}_{E=E_{min}} = 0,
    \label{eq:1DFP_BCE}                                                          \\
    \text{BCS: } & \varphi_{FP}(0, E) = A e^{-\qty(\frac{E - E_0}{\sigma_E})^2},
    \label{eq:1DFP_BCS}
\end{align}
to ensure a unique solution. The energy boundary conditions are of the Dirichlet and Neumann type, while the space boundary condition is a Gaussian function with amplitude $A$, nominal beam energy $E_0$ and spread $\sigma_E$. Gerbershagen \citep{gerbershagenSimulationsMeasurementsProton2017} showed that this is a realistic energy spectrum for protons that underwent energy degradation. After discretizing the system in energy a so called semi-discrete system of equations is obtained that is thereafter solved using the SDIRK3 method\footnote{When compared to the Crank-Nicholson method in our earlier work \citep{burlacuDeterministicAdjointBasedSemiAnalytical2023a}, this method increased the accuracy of the Fokker-Planck fluxes without degrading the speed of the algorithm.}.

\subsubsection{The Fermi-Eyges equation}

The advantage of the Fermi-Eyges equation \ref{eq:FE} is its analytical solution via Fourier transforms \citep{gebackAnalyticalSolutionsPencilBeam2012}, namely
\begin{align}
    \label{eq:FE_sol}
    \varphi_{FE}(z, \vb*{\rho}, \vu*{\Omega}) = \frac{A^2}{4 \pi^2}
    \frac{\exp\qty(-\frac{|\vb*{\rho}|^2}{2 \overline{\xi^2}(z)})}
    {\overline{\xi^2}(z)} \frac{\exp\qty(-\frac{1}{2 B(z)} \qty|
        \vu*{\Omega} - \frac{\overline{\theta\xi}(z)}
        {\overline{\xi^2}(z)}\vb*{\rho}|^2 )}{B(z)}.
\end{align}
The solution from Equation \ref{eq:FE_sol} is a Gaussian in the beam lateral coordinates $\vb*{\rho} = (x, y) $ and in the angular coordinates $\vu*{\Omega} = (\Omega_x, \Omega_y)$ with its depth-dependent FE coefficients, namely $\overline{\theta^2}(z)$ (variance of the angular direction), $\overline{\xi^2}(z)$ (variance of the lateral position), $\overline{\theta\xi}(z)$ (covariance of the lateral position and angular direction) \citep{gottschalkTechniquesProtonRadiotherapy2012}, determined by the material path encountered along the central axis of the beam. The solution $\varphi_{FE}$ is obtained by imposing a boundary condition that is a product of two identical double Gaussians, one in $(x, \Omega_x)$ and one in $(y, \Omega_y)$.

\begin{figure}[!h]
    \centering
    \resizebox{.6\textwidth}{!}{
        \input{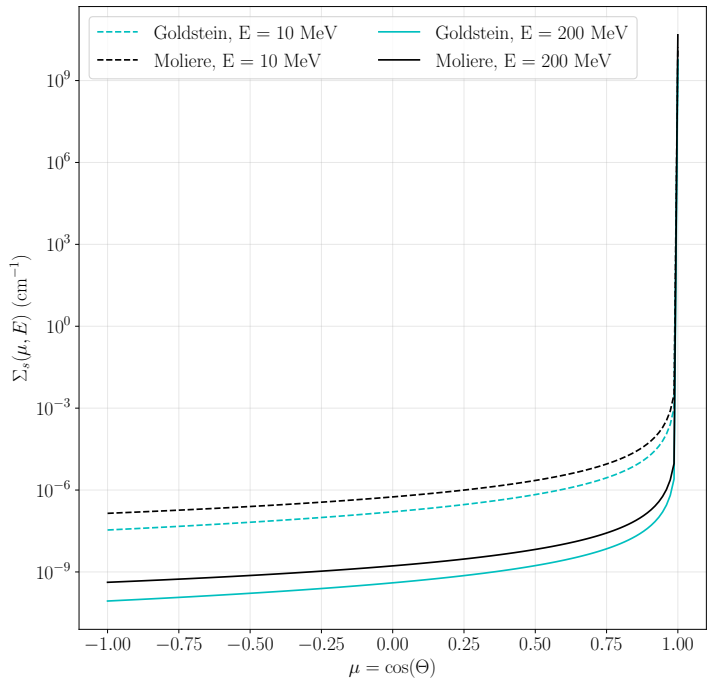}
    }
    \caption{Comparison of different macroscopic elastic scatter cross section models for protons in water.}
    \label{fig:Sigma_e_comparison}
\end{figure}

As shown in Subsection \ref{subsec:metric_definition}, to obtain the dose in a physical region only the $\overline{\xi^2}$ coefficient is needed. This is the second moment of $\overline{\Sigma_{tr}}$ and is computed as
\begin{equation}
    \overline{\xi^2}(z)      = \overline{\xi^2}(0)
    + 2 \overline{\theta \xi}(0) z
    + \overline{\theta^2}(0) z^2
    +  \int\limits_{0}^z (z-z')^2 \overline{\Sigma_{tr}}(z') \dd z'
\end{equation}
with $\overline{\xi^2}(0)$, $\overline{\theta^2}(0)$ and $\overline{\theta \xi}(0)$ constants based on the imposed double Gaussian boundary condition. The quantity $\overline{\Sigma_{tr}}$ is the depth-dependent energy spectrum (i.e., $\varphi_{FP}$) averaged\footnote{In the original formalism, $\Sigma_{tr}$ depends on the average depth-dependent beam energy $E_a(z)$. It was found that weighing $\Sigma_{tr}$ with the depth-dependent energy spectrum yields more accurate lateral profiles that better match MC results.} macroscopic transport cross section $\Sigma_{tr}$, namely
\begin{align*}
    \overline{\Sigma_{tr}}(z) & = \int \dd E
    \varphi_{FP}(z, E) \Sigma_{tr}(z, E) \bigg/ \int \dd E \varphi_{FP}(z, E),
\end{align*}
with the macroscopic transport cross $\Sigma_{tr}$ computed using the  macroscopic elastic scatter cross section $\Sigma_s$ via
\begin{align*}
    \Sigma_{tr}(z, E) = \int\limits_{-1}^{1} \dd \mu
    \Sigma_s(z, E, \mu) \qty(1-\mu)  \text{, with }
    \mu=\cos(\vu*{\Omega} \cdot \vu*{\Omega}').
\end{align*}
There are multiple elastic scatter models that can be used to compute $\Sigma_s$. In this work, the two models that were investigated were the small-angle first Born approximation scatter model and Moliere's model \citep{scottTheorySmallAngleMultiple1963}. Moliere's model provides improvements over the first Born approximation as it is valid for large angles, does not assume that the nucleus has infinite mass and includes the contribution of electronic screening of the nucleus. A comparison between these two models is shown in Figure \ref{fig:Sigma_e_comparison} where it can be seen that Moliere's model predicts an increased macroscopic elastic scatter cross section for all energies. This implies a larger transport cross section $\Sigma_{tr}$ which in turn implies a larger variance of the lateral position of the FE solution that better matches the lateral profiles obtained from TOPAS.

\subsection{Metric definition}
\label{subsec:metric_definition}

The 6-dimensional phase-space density resulting from the individual solutions to the FP and FE equations can be used to obtain all clinically relevant metrics. For example, let $\Psi_{FE}$ be the angular integral of $\varphi_{FE}$, namely
\begin{align}
    \Psi_{FE}(\vb*{r})
    = \int\limits_{4 \pi} \dd \vu*{\Omega} \varphi_{FE}(\vu*{\Omega}, \vb*{r})
    = \frac{A^2}{2 \pi \overline{\xi^2}(z)} \exp(-\frac{x^2 + y^2}{ 2 \overline{\xi^2}(z) }),
\end{align}
and let $\Psi_{FP}$ be
\begin{align}
    \label{eq:Psi_FP}
    \Psi_{FP}(z)
     & = \int\limits_{0}^\infty \dd E E \qty[-
        \pdv{S(z, E) \varphi_{FP}}{E}
        - \frac{1}{2} \pdv[2]{T(z, E) \varphi_{FP}}{E}
    + \Sigma_a(z, E) \varphi_{FP}].
\end{align}
Then, if the CT image volume is given by the union of all of its $N_v$ voxels (i.e., $\mathscr{V} = \bigcup V_k, k=1, \ldots, N_v$ where $V_k$ is the volume of one voxel), the energy $E_k$ deposited by the proton beam in a voxel $V_k$ is given by
\begin{align*}
    E_k = \int_{V_k} \dd V \Psi_{FE}(\vb{r}) \Psi_{FP} (z).
\end{align*}
The dose $D_k$ in the same voxel $k$ is given as
\begin{align*}
    D_k = \frac{E_k}{m_k} = \frac{1}{\Delta V} \int_{V_k} \dd V \frac{\Psi_{FE}(\vb{r}) \Psi_{FP} (z)}{\rho_k},
\end{align*}
where $\Delta V = \Delta x \Delta y \Delta z$ is the volume of a voxel $k$ (constant for all voxels in the CT image) and $\rho_k$ is the mass density of voxel $k$. Thus, the total dose in a certain region of interest (ROI) of the CT image, identified by the union of its corresponding voxels, is the sum of $D_k$ over all $k$ in the ROI. 

\subsection{Accounting for lateral heterogeneities}

As equations \ref{eq:1DFP} and \ref{eq:FE} show, the formalism presented is inherently unable to account for heterogeneities located laterally to the central beam axis. To account for such cases, two modifications are introduced. First, the deposited energy density is multiplied by a laterally-dependent density scaling. This is physically motivated as the deposited energy is directly proportional to the local density. Second, each treatment plan spot (i.e.,  pencil beam) is decomposed into several sub-spots (i.e., beamlets) that are placed on concentric rings arround the original spot position. The properties of the rings and of the beamlets on the rings are optimized for best performance.

\subsubsection{Lateral density scaling}

The energy density in a voxel $k$ is scaled by the ratio of the density $\rho_{ck}$ on the central beam axis at a depth that corresponds to the voxel $k$ and the density $\rho_k$ of the voxel itself, namely 
\begin{align}
    E_{k} & = \int\limits_{V_k} \dd V \frac{\rho_k}{\rho_{ck}} \Psi_{FE}(\vb{r}) D_{FP}(z) \label{eq:E_dep_def},
\end{align}
Using this scaling, the dose in voxel $k$ becomes
\begin{align}
    D_{k} = \frac{1}{\Delta V}\int\limits_{V_k} \dd V 
    \frac{\Psi_{FE}(\vb{r}) D_{FP}(z)}{\rho_{ck}}  \label{eq:D_dep_def}.
\end{align}
Thus, a pencil beam distributes laterally a dose proportional to the density along the central beam axis. 

\subsubsection{Optimized Gaussian beam splitting}

\begin{figure}[H]
    \centering
    \begin{subfigure}{0.49\textwidth}
        \includegraphics[width=\columnwidth]{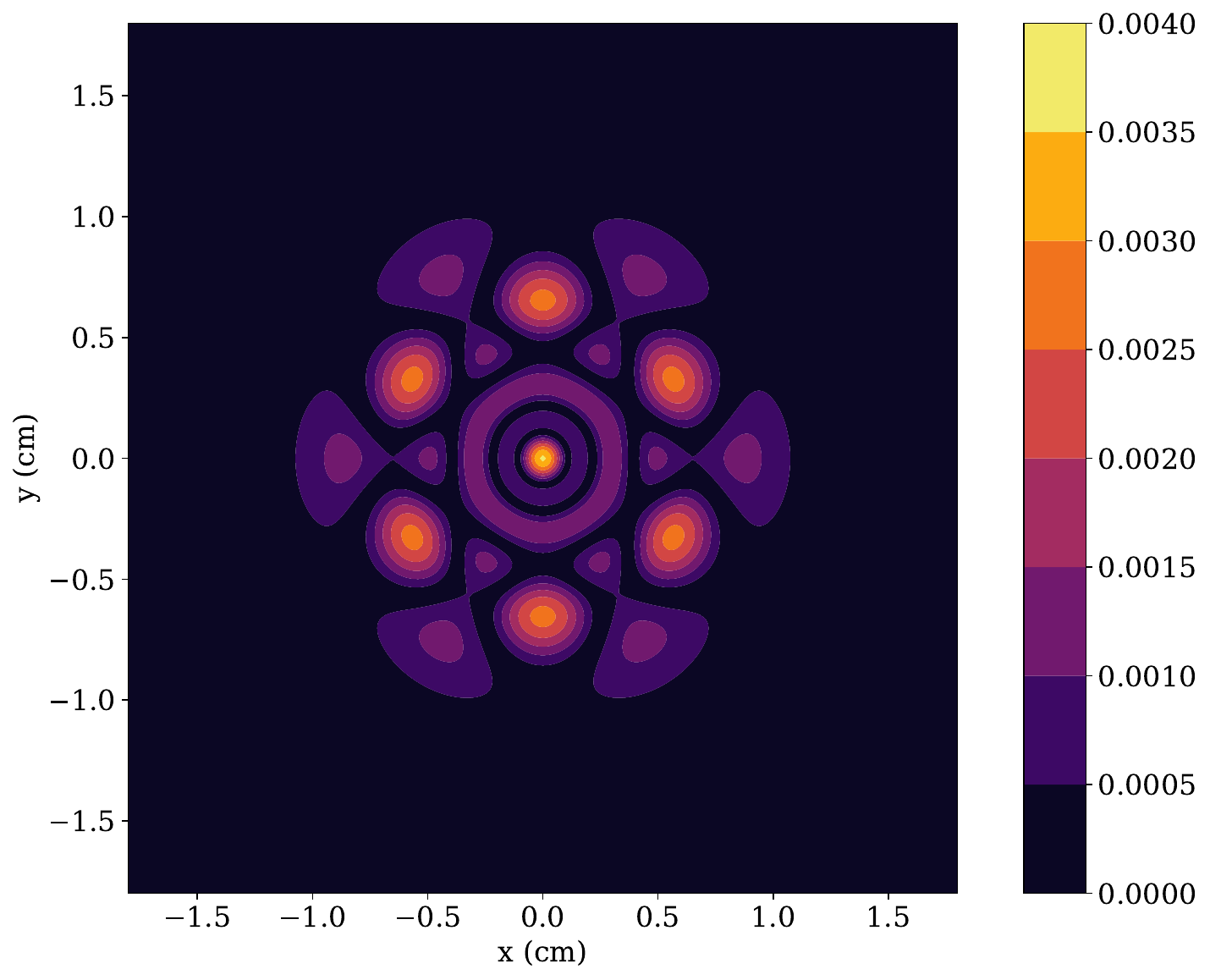}
        \caption{Configuration 1+6+6}
    \end{subfigure}
    \hfill
    \begin{subfigure}{0.4\textwidth}
        \includegraphics[width=\columnwidth]{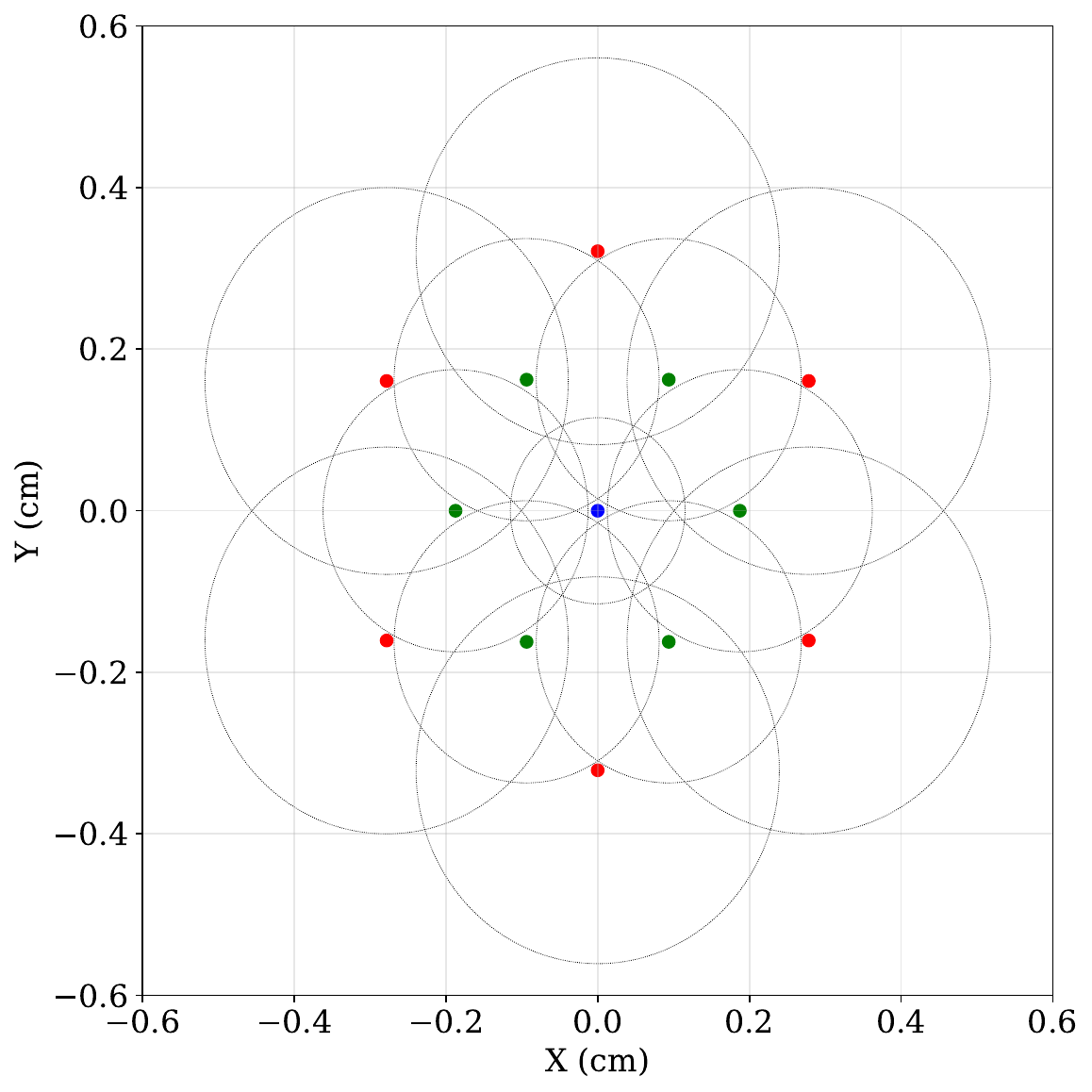}
        \caption{Configuration 1+6+6}
    \end{subfigure}
    \begin{subfigure}{0.49\textwidth}
        \includegraphics[width=\columnwidth]{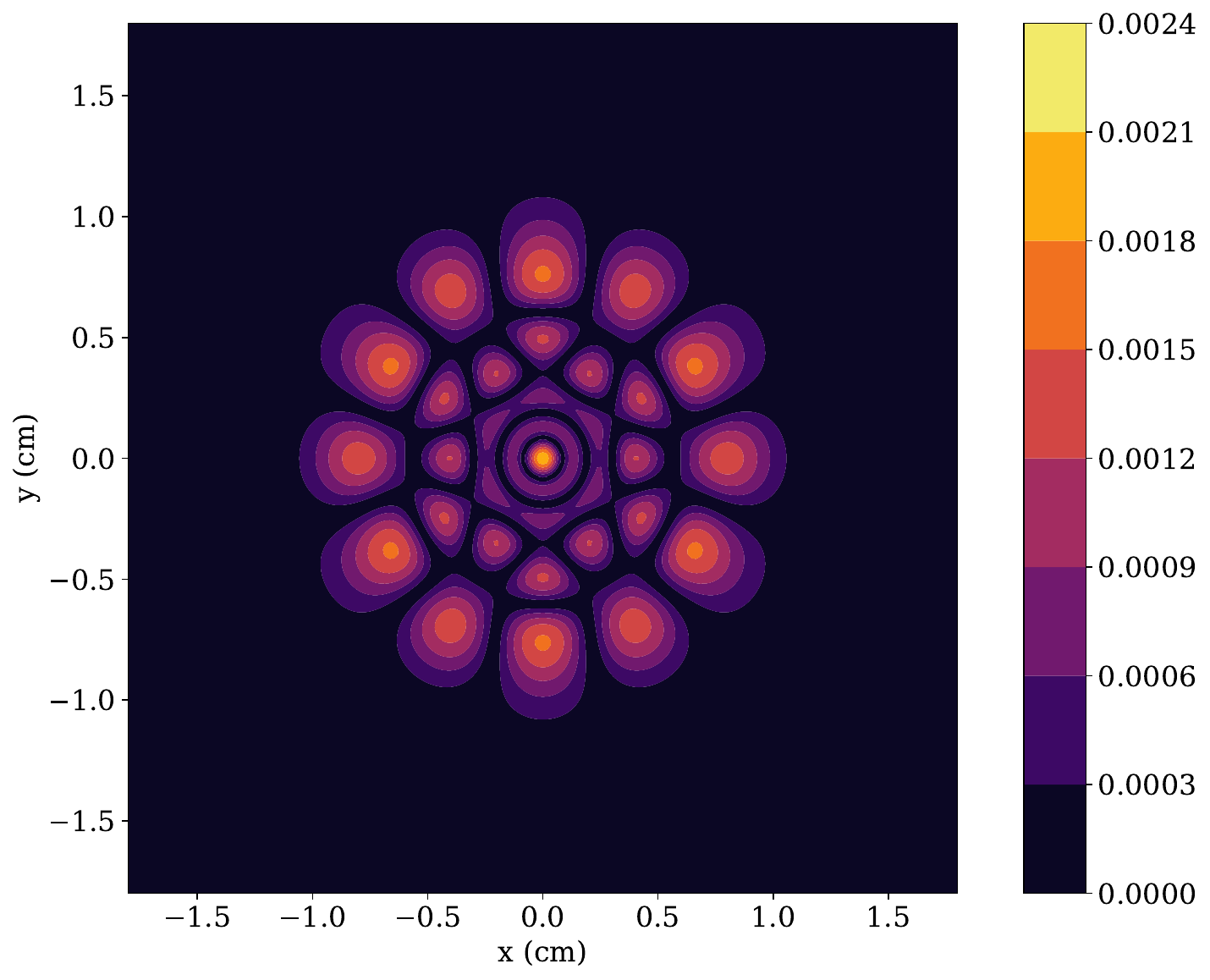}
        \caption{Configuration 1+6+6+6}
    \end{subfigure}
    \hfill
    \begin{subfigure}{0.4\textwidth}
        \includegraphics[width=\columnwidth]{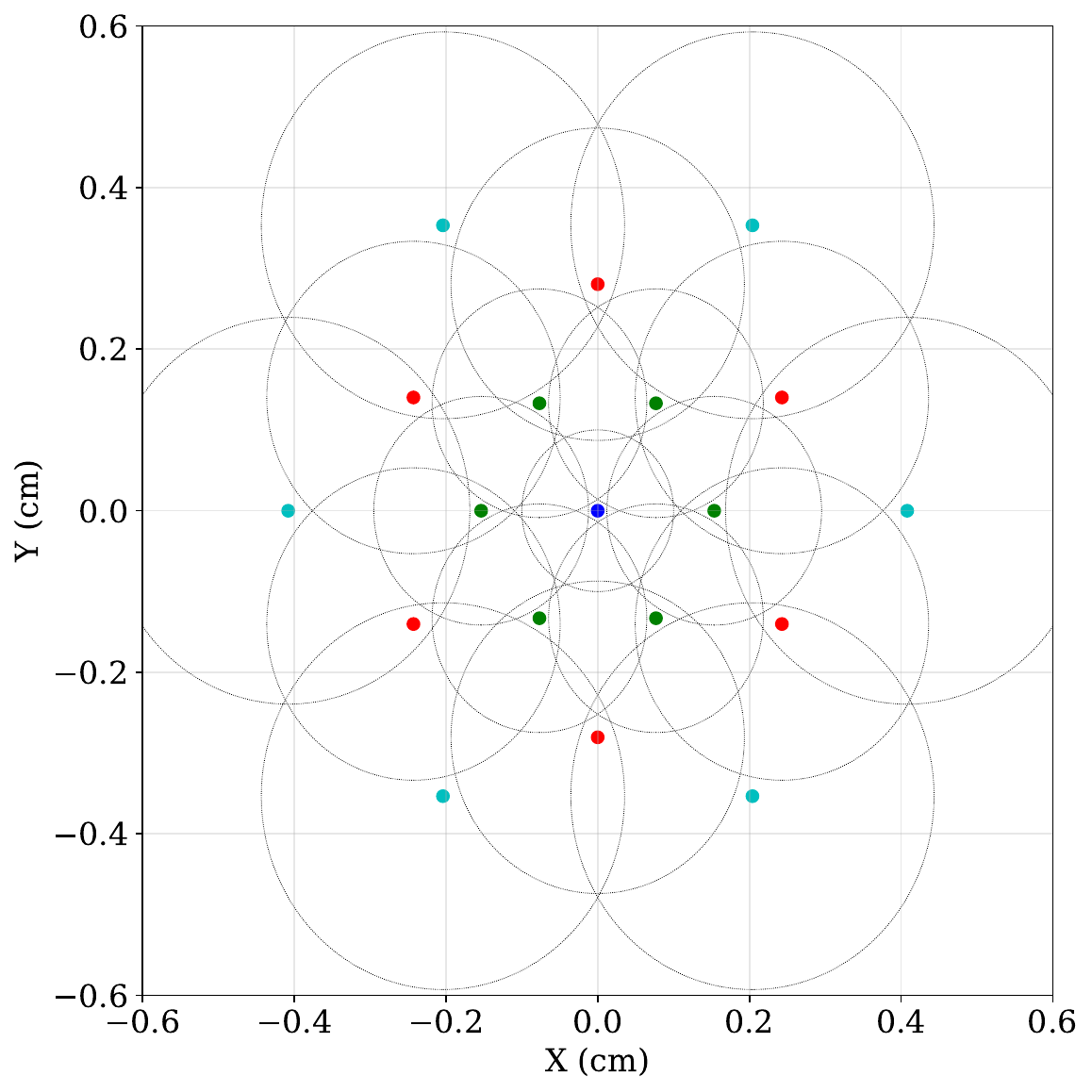}
        \caption{Configuration 1+6+6+6}
    \end{subfigure}
    \hfill
    \begin{subfigure}{0.49\textwidth}
        \includegraphics[width=\columnwidth]{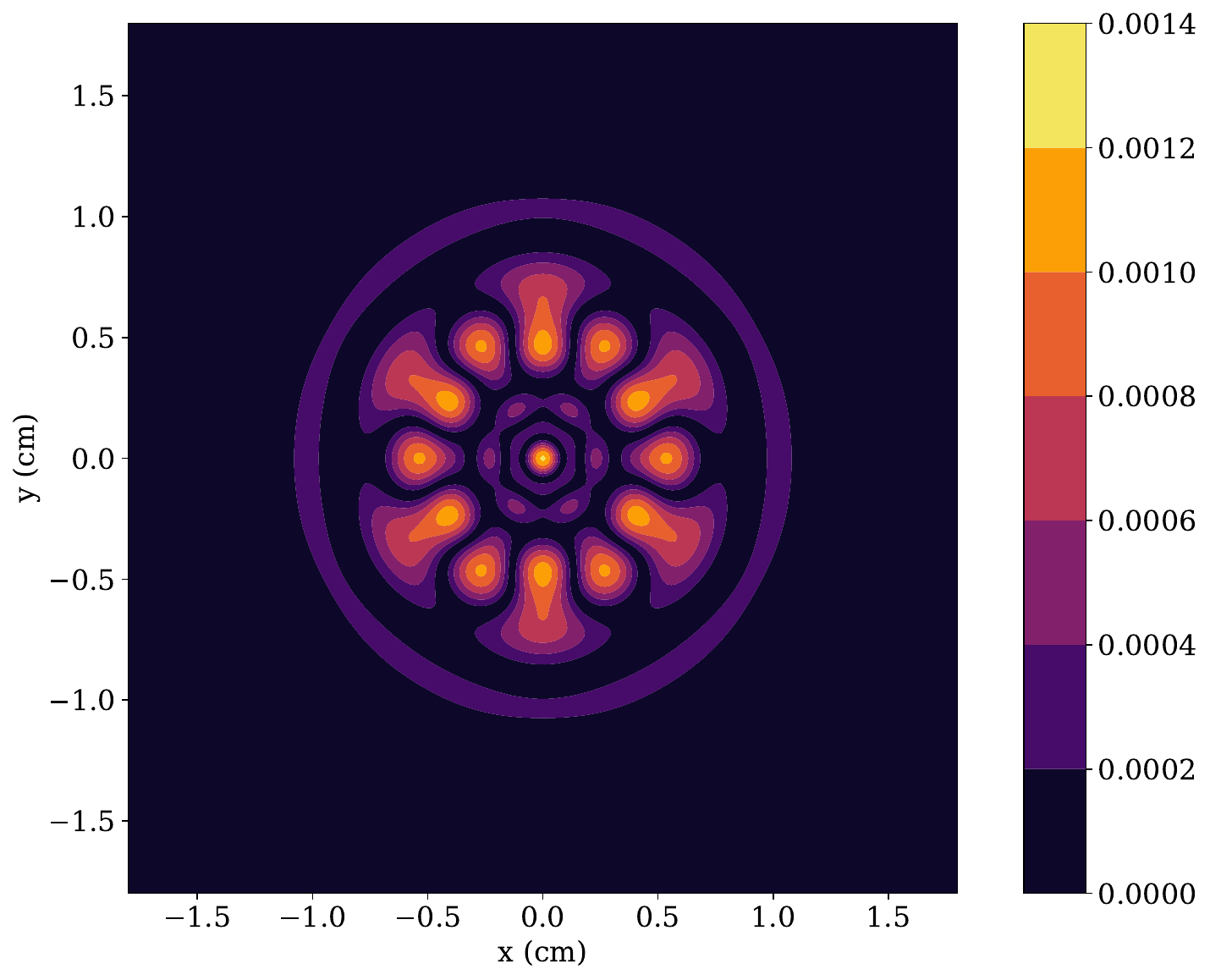}
        \caption{Configuration 1+6+6+12}
    \end{subfigure}
    \hfill
    \begin{subfigure}{0.4\textwidth}
        \includegraphics[width=\columnwidth]{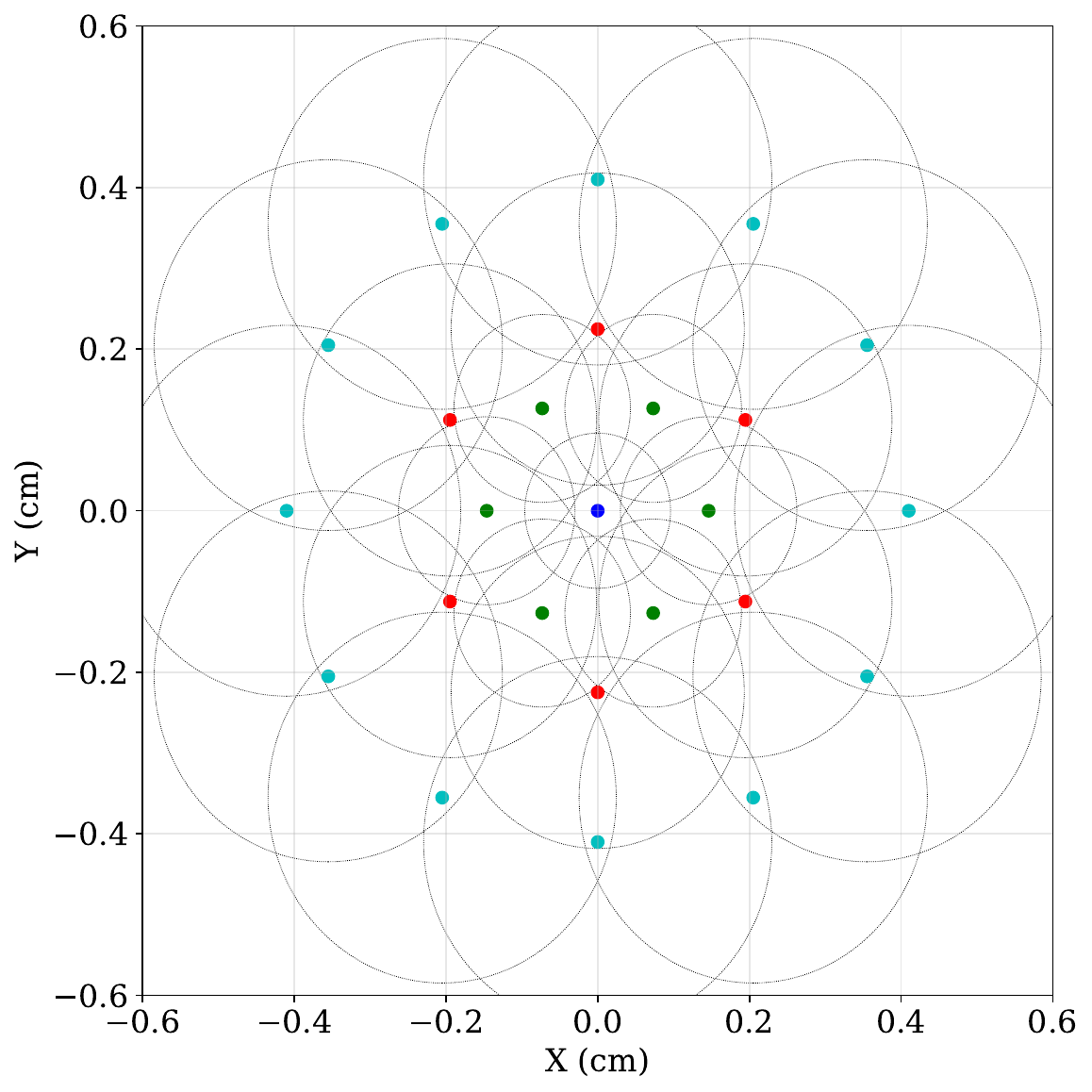}
        \caption{Configuration 1+6+6+12}
    \end{subfigure}
    \hfill
    \caption{Left column: absolute normalized difference between $\Psi_{FE}^0$ and $\Psi_{FE}^a$ using the optimized parameters. Right column: corresponding physical optimized positions of the individual beamlets in the lateral plane. Points with the same color are on the same ring, circle radii are $\sigma_i$ on ring $i$.}
    \label{fig:gaussian_splitting}
\end{figure}

On the boundary of the computational domain, the lateral dependence of the six-dimensional phase-space density is described by
\begin{align}
    \label{eq:phi_FE_0}
    \Psi_{FE}^{z=0}(x,y) =
    \int\limits_{4 \pi} \varphi_{FE}(x,y,z=0,\Omega_x,\Omega_y)
    \dd \vu*{\Omega}
    = \frac{1}{2 \pi \sigma_s^2}
    \exp\qty(-\frac{(x^2 + y^2)}{2 \sigma_s^2}),
\end{align}
where $\sigma_s$ is the spatial standard deviation or spread of the x and y symmetric Gaussian. For the purpose of lateral beam splitting the original spot's central axis is placed at the origin of a 2D lateral grid. Given the radial symmetry of the Gaussian, placing sub-spots or beamlets on $N_r + 1$ concentric rings with radii $r_i$ around the original spot location was chosen, in a similar manner to Yang's method \citep{yangImprovedBeamSplitting2020}. On a given ring $i$ the beamlets share the same weight $w_i$ and spread $\sigma_i$. The zeroth ring has a radius equal to zero and a single beamlet that is placed at the origin of the 2D lateral grid. Thus, the approximated fluence $\Psi_{FE}^a$ is written as
\begin{align}
    \label{eq:phi_FE_approx}
    \Psi_{FE}^{a}(x,y) & = \displaystyle\sum_{i=0}^{N_r}
    \displaystyle\sum_{k=1}^{n_i}
    \frac{w_i}{2 \pi \sigma_i^2}
    \exp\qty(-\frac{(x - x_{ik})^2 + (y - y_{ik})^2}{2 \sigma_i^2}), \\
    x_{ik}             & = r_i \cos(\frac{2 \pi k}{n_i} + \alpha_i),
    y_{ik} = r_i \sin(\frac{2 \pi k}{n_i} + \alpha_i), \nonumber
\end{align}
with $n_i$ being the number of sub-spots placed on ring $i$, $(x_{ik}, y_{ik})$ are the coordinates of a sub-spot with index $k$ on ring $i$ and $\alpha_i$ is a ring-dependent angular offset (meant to improve coverage for consecutive rings with the same number of beamlets). Prior to the optimization the number of rings $N_r$, the number of points on each ring $n_i$ and the ring offsets $\alpha_i$ are specified. As opposed to Yang's \citep{yangImprovedBeamSplitting2020} approach this formalism and implementation is not restricted to a number of pre-defined schemes. In principle any number of beamlets per ring and number of rings can be optimized. The optimization parameters (weights, spreads and ring radii) are collected in a vector denoted by $\vb*{\theta} \in \mathbb{R}^{3(N_r+1)}$ with a structure of $\vb*{\theta} = (\ldots, w_i, r_i, \sigma_i, \ldots)$. The objective function of the optimization problem is defined as
\begin{align*}
    J(\vb*{\theta}) = \iint\limits_{-10\sigma_s}^{10\sigma_s} \dd x \dd y \,(\Psi_{FE}^a - \Psi_{FE}^{z=0})^2 \bigg/ \iint\limits_{-10\sigma_s}^{10\sigma_s} \dd x \dd y (\Psi_{FE}^{z=0})^2,
\end{align*}
and is input into a scipy implementation of a trust-region constrained algorithm \citep{virtanenSciPyFundamentalAlgorithms2020, laleeImplementationAlgorithmLargeScale1998}. The weights $w_i$ are bound constrained to be in the unit interval, namely $0 \leq w_i \leq 1, \forall i=0, \ldots, N_r$ and are constrained such that
\begin{equation*}
    \displaystyle\sum_{i=0}^{N_r} w_i n_i = 1,
\end{equation*}
in order to ensure particle number conservation. To further guide the highly degenerate solution space towards useful splitting schemes, the ring radii are bound according to the initial spatial spread of the 2D Gaussian $\sigma_s$ such that $0 \leq r_i \leq r_{i+1} \leq 2 \sigma_s$. This evenly distributes the rings in $[0, 2 \sigma_s]$ and avoids optimal but less useful configurations where all the rings are placed close to one another and the origin. Similarly, the spreads of the rings $\sigma_i$ are bound such that $0.3 \sigma_s \leq \sigma_i \leq \sigma_{i+1} \leq 0.8 \sigma_s$. The first ring should have the smallest spread so that errors coming from the central axis are limited. In the case of a spot with an initial spread of $\sigma_s = \SI{0.3}{\cm}$ Figure \ref{fig:gaussian_splitting} shows for three different splitting schemes the absolute difference between $\Psi_{FE}^{z=0}(x,y)$ and $\Psi_{FE}^{a}(x,y)$ in the left column and the actual positions of the beamlets on the concentric rings together with the optimized spreads (indicated by the circle radii) around each spot in the right column.

\subsection{Metric change computation}

Next to its dose computation capabilities, an advantage of YODA is the ease of applying the adjoint method. This general mathematical framework approximates to first order the change in a metric as a function of the change in all independent variables. Examples of possible independent variables are HU values in the CT image or treatment plan spot characteristics such as mean energy, energy spread, position, MU value (or equivalently the number of protons), angular spread and the spot size. Examples of metrics are the mean dose to an OAR or NTCP values. The adjoint method is useful when the number of independent variables is large (so that re-computing the metric for each new variable becomes prohibitively expensive) and their change is relatively small (so that the first order adjoint approximation is accurate). Examples of applications are computing dose or NTCP differences caused by differences between planned and delivered spot MU values or isocenter positions or by delivering yesterday's treatment plan on today's CT image. Since CTs typically have millions of voxels this is likely always the case in radiotherapy. This section provides only the the main details of the adjoint method for the case when the independent variables that change are the HU values of the CT image and the metric considered is the dose in a voxel $V_k$. Larger regions of clinical interest are trivial generalizations of this case. 

A given change in the HU values of the CT image implies two distinct changes in the deposited dose $D_k$ in the voxel $k$. One is a direct change, since a HU change in the voxel $k$ implies, among others, a stopping power change which can be directly inputted in the $D_k$ change via Equation \ref{eq:Psi_FP}. The other is an indirect change, as a stopping power change somwhere along the proton beam path implies a proton flux change in the considered voxel $k$. This change can only be known by re-solving for $\varphi$ from the FP and FE equations with the new HU values. Thus, the change in $D_k$ is written as,
\begin{align*}
    \var{D_k} = \var{D_{k, dir}} + \var{D_{k, indir}},
\end{align*}
where $\delta$ denotes a variation, $\var{D_{k, dir}}$ denotes the part of $\var{D_k}$ that can be directly computed and $\var{D_{k, indir}}$ denotes the part that would have to be re-computed.

The adjoint method removes from $\var{D_k}$ the part $\var{D_{k, indir}}$ that would have to be re-computed and in this process computes a first order approximation to $\var{D_k}$. This is done by expressing $\var{D_{k, indir}} $ as an inner product between two quantities. One is the change in the proton flux $h_\varphi$ caused by the change in the HU values and the other is a vector denoted by $r^\dagger$, namely $\var{D_{k, indir}} = \langle h_\varphi, r^\dagger \rangle $. The vector $r^\dagger$ is identified as the right-hand side of a new system called the adjoint system. This system is written as $L^\dagger \varphi^\dagger = r^\dagger$ and its solution is called the adjoint flux $\varphi^\dagger$. Using this together with the properties of the adjoint system $\var{D_{k, indir}} $ is expressed as
\begin{align}
    \label{eq:indir_effect}
    \var{D_{k, indir}} & = \langle h_\varphi, r^\dagger \rangle = \langle h_\varphi, L^\dagger \varphi^\dagger \rangle = \langle L h_\varphi, \varphi^\dagger \rangle = - \langle \var{L} \varphi_{FP}, \varphi^\dagger \rangle \\
                       & = \iint \dd E \dd z \varphi^\dagger(z, E) \qty[ - \pdv{\varphi_{FP}}{z} + \pdv{\var{S} \varphi_{FP}}{E} + \pdv[2]{\var{T}\varphi_{FP}}{E} - \var{E_a} \varphi_{FP}], \nonumber
\end{align}
where $\var{S}, \var{T}, \var{\Sigma_a}$ are the changes in the stopping power, straggling coefficient and macroscopic absorption cross section caused by the change in the voxel HU value. Thus, if there are $N_{HU}$ values for which the dose in the voxel $k$ is desired, re-computing would cost $N_{HU}$ FE and FP solutions. In contrast, the adjoint method only performs two FE and FP solutions and $N_{HU}$ inner products. The construction of the right hand side $r^\dagger$ and of the adjoint operator $L^\dagger$ is illustrated in our previous work \citep{burlacuDeterministicAdjointBasedSemiAnalytical2023a}. This approach can be advantageous for the time constrained cases when the changes in the CT image HU values are small enough. For such cases, the adjoint method can provide significant time savings by avoiding an expensive re-computation of the treatment plan on the new CT image.

\section{Results and discussion}
\label{sec:results_and_discussion}

\subsection{Dose engine performance}

The dose engine of YODA was benchmarked against TOPAS in homogeneous and hetereogeneous water tanks, H\&N, prostate and lung CTs. TOPAS simulations were performed using the em-opt4 physics list which is the most accurate modelling of electromagnetic interactions available within TOPAS. Nuclear interactions were excluded from this comparison as YODA does not currently account for nuclear interactions. In all TOPAS simulations the number of protons per spot was set to \num{1.0e8} and the maximum number of available cores (\num{48}) was used. Using this physics list and number of cores, the run-times of TOPAS was in the order of hours. In all test cases, a YODA spot was split according to a $1+6+6+12+12+24$ Gaussian beam splitting scheme as this was found to yield accurate results when compared to TOPAS. For this splitting scheme on average one spot takes \SI{2}{\second} to compute. Additional speed-ups could be achieved in two ways. One is to address the main speed limitation (memory access bandwidth) by implementing the algorithm on a graphics processing unit card. The second is to implement an adaptive energy grid on a per sub-spot level. Currently the energy grid is divided into a fixed number of groups which results in the majority of the groups and thereby the system solved at each step being empty. By adapting the energy grid to be finely discretized in the locations in energy where the flux has significant values and coarse everywhere else significant speed-ups can be expected.


\subsubsection*{Simplified tank geometries}

\begin{figure}[H]
    \centering
    \begin{subfigure}{\linewidth}
        \centering
        \includegraphics[width=0.9\linewidth, trim={0 7.1cm 0 7.6cm}, clip]{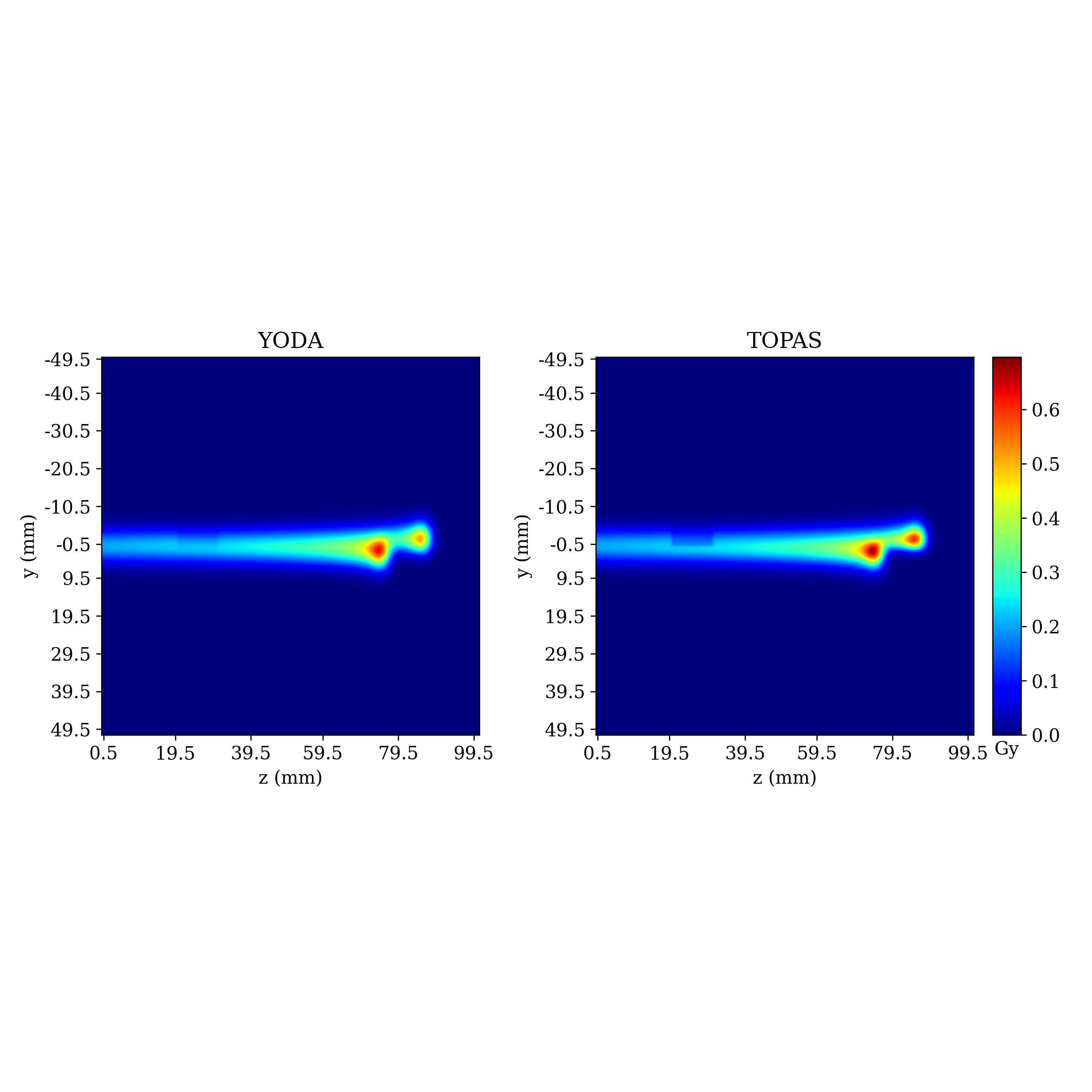}
        \caption{-1000 HU}
    \end{subfigure}
    \begin{subfigure}{\linewidth}
        \centering
        \includegraphics[width=0.9\linewidth, trim={0 7.1cm 0 7.6cm}, clip]{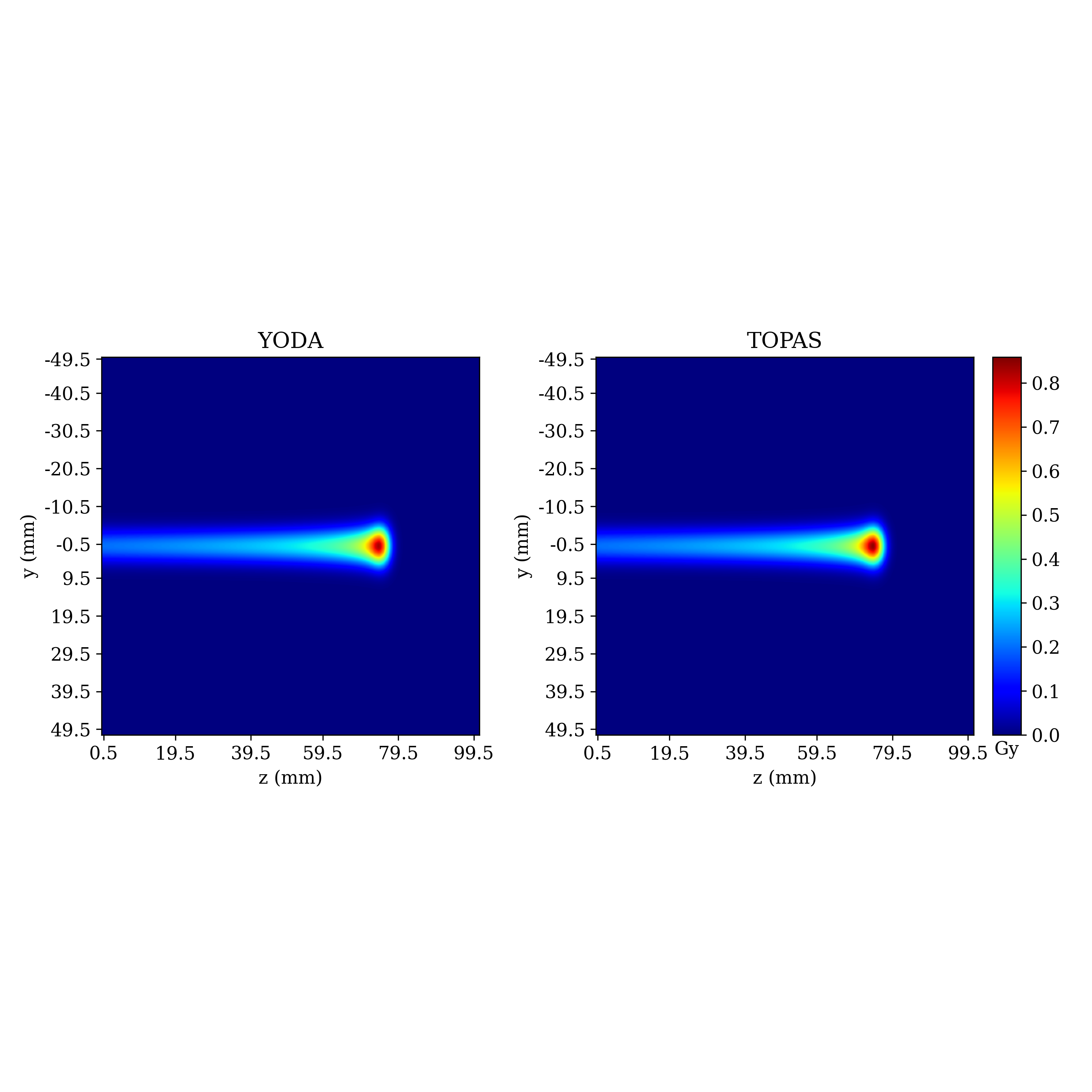}
        \caption{0 HU}
    \end{subfigure}
    \begin{subfigure}{\linewidth}
        \centering
        \includegraphics[width=0.9\linewidth, trim={0 7.1cm 0 7.6cm}, clip]{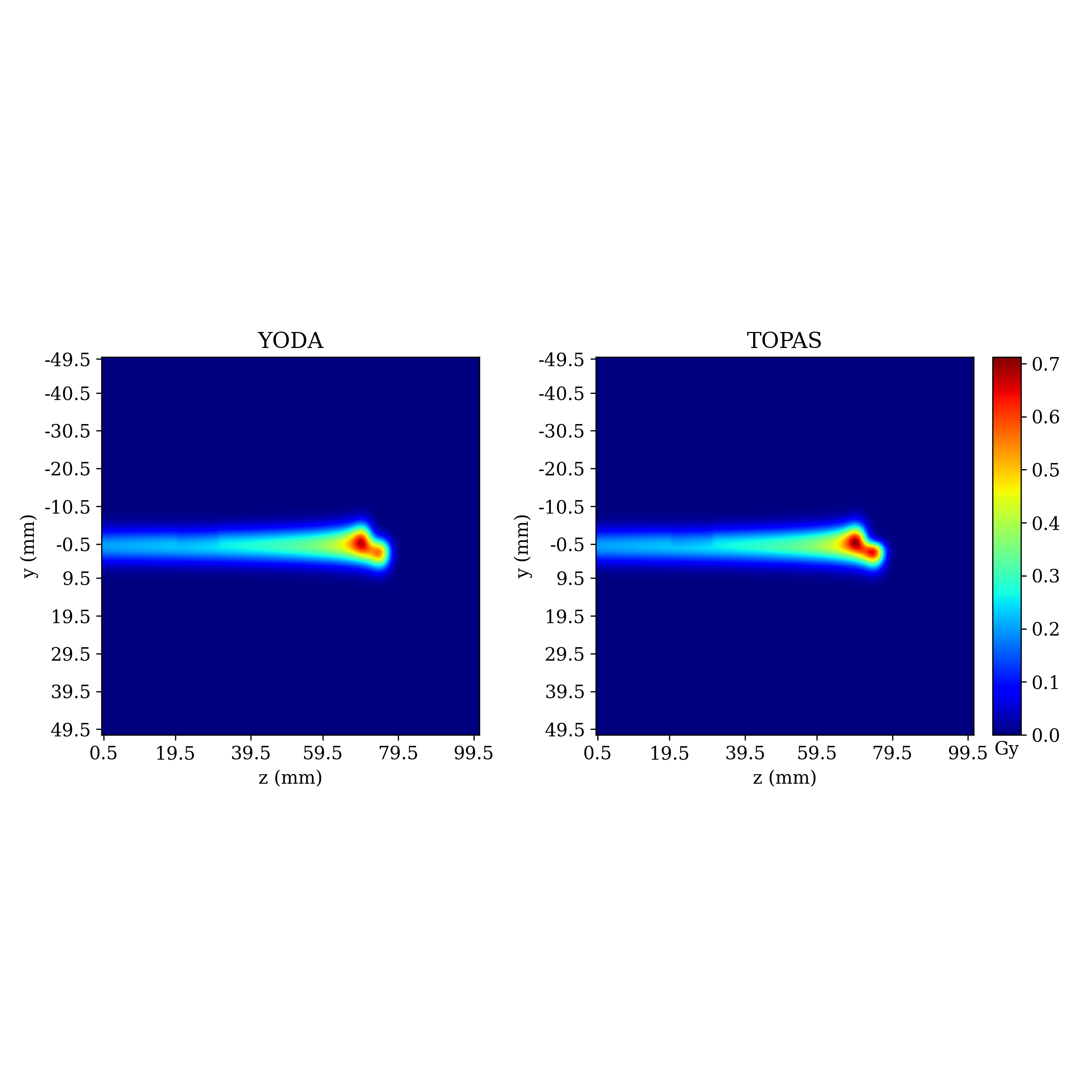}
        \caption{+1000 HU}
    \end{subfigure}
    \caption{2D dose slices for YODA and TOPAS in the simplified tank geometries.}
    \label{fig:2d_plots_tank}
\end{figure}

\begin{figure}[H]
    \centering
    \begin{subfigure}{\linewidth}
        \centering
        \includegraphics[width=0.7\linewidth]{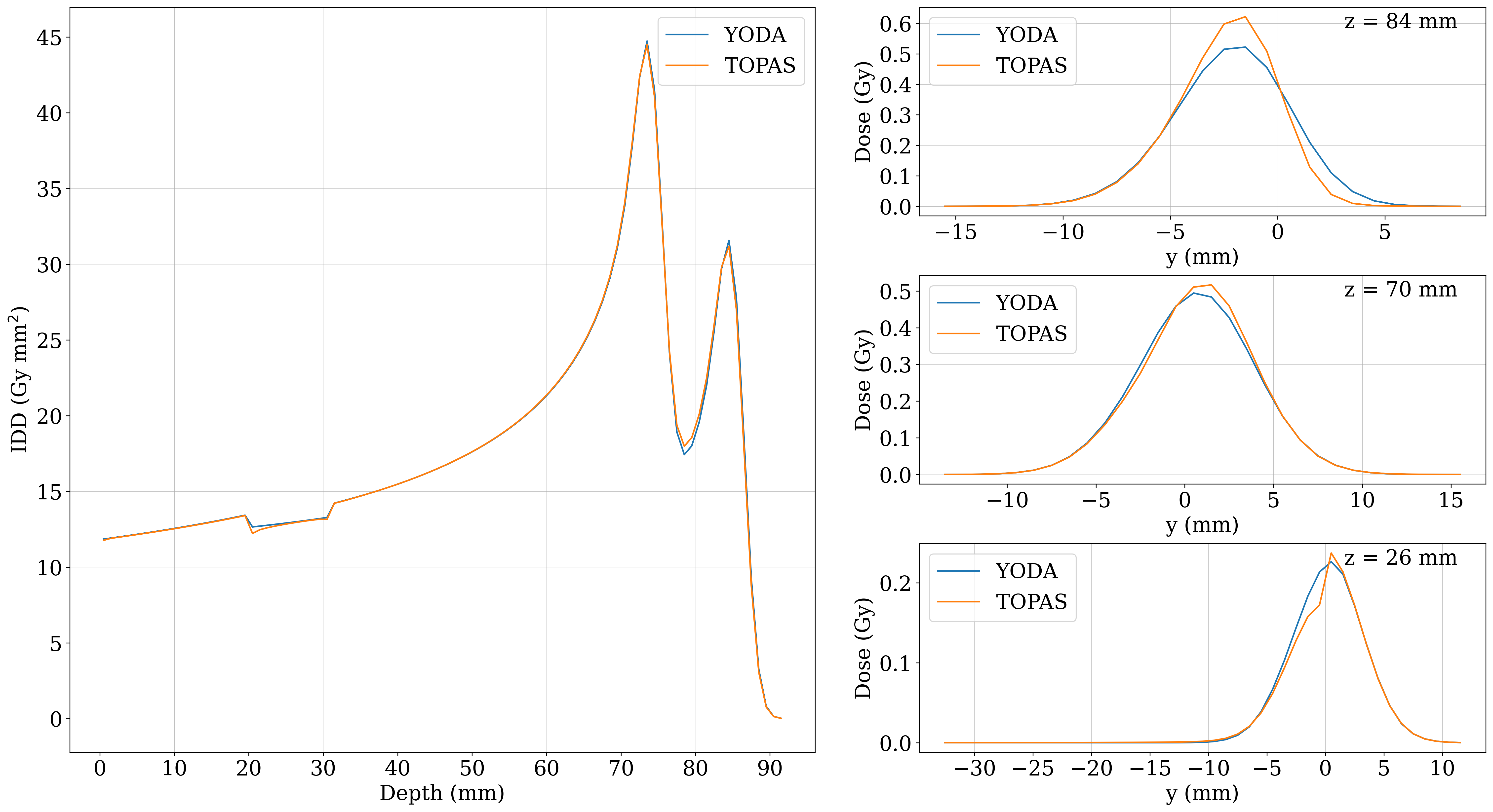}
        \caption{-1000 HU}
    \end{subfigure}
    \begin{subfigure}{\linewidth}
        \centering
        \includegraphics[width=0.7\linewidth]{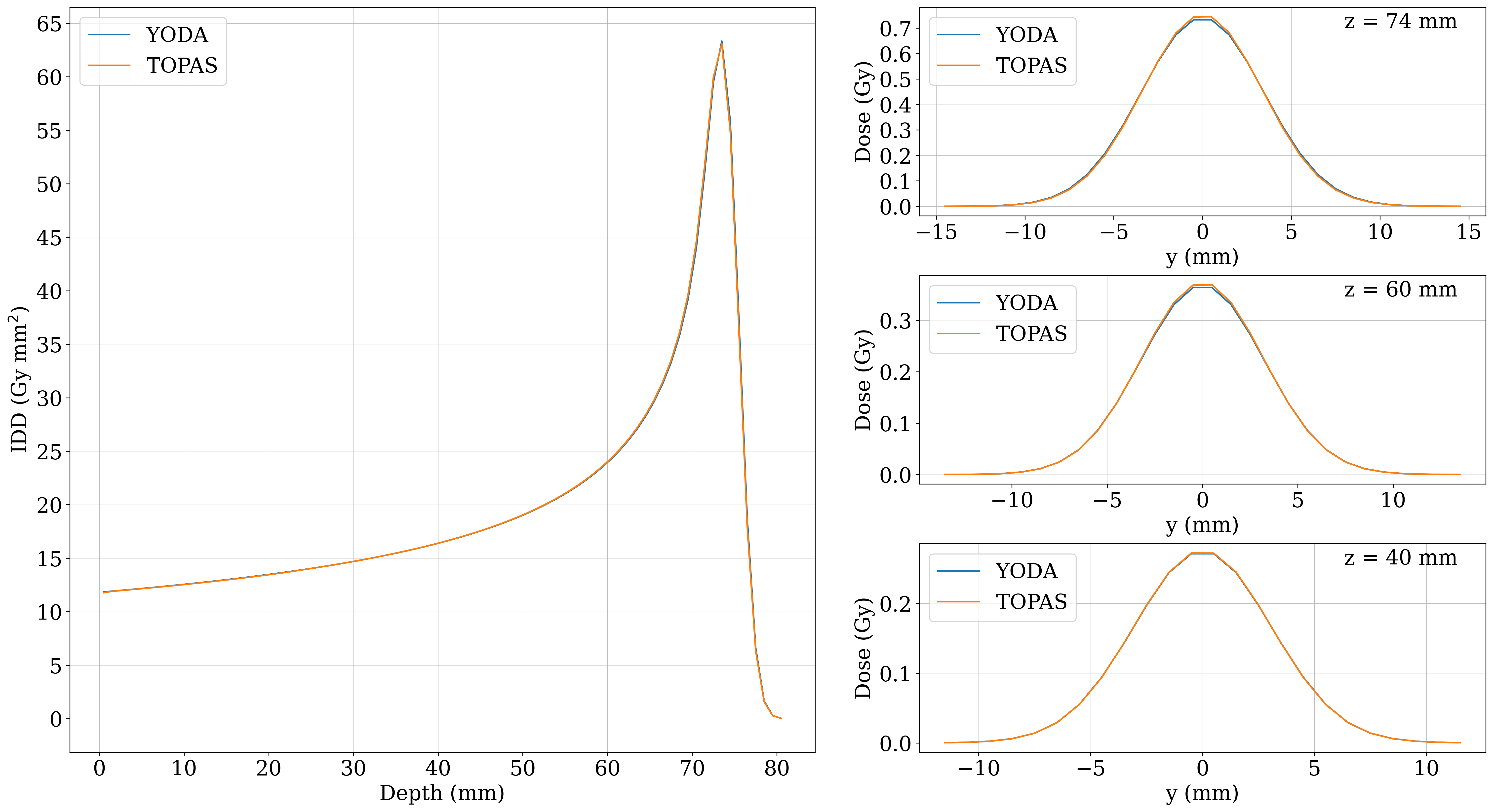}
        \caption{0 HU}
    \end{subfigure}
    \begin{subfigure}{\linewidth}
        \centering
        \includegraphics[width=0.7\linewidth]{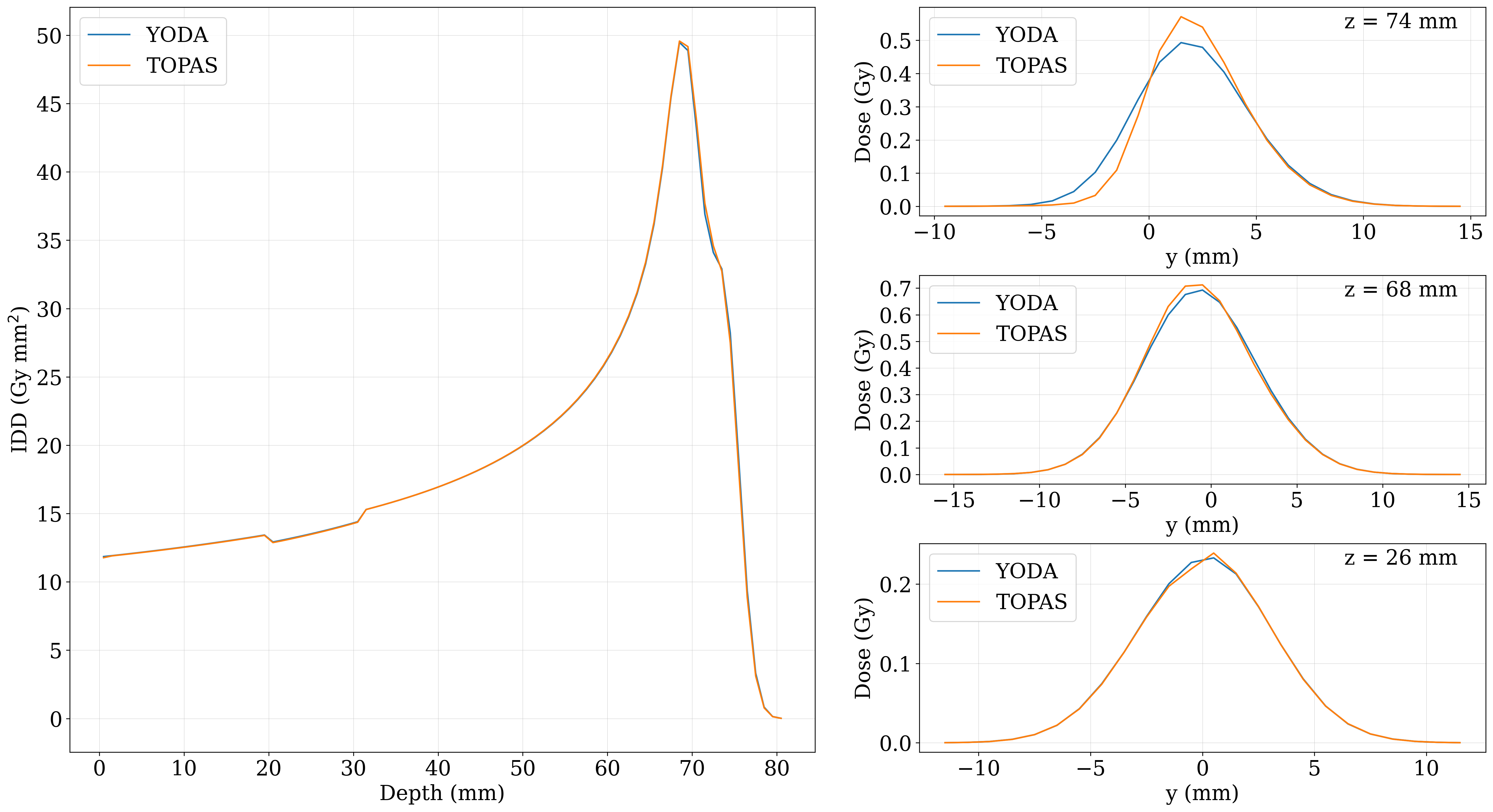}
        \caption{+1000 HU}
    \end{subfigure}
    \caption{IDDs and lateral profiles for YODA and TOPAS in the simplified tank geometries.}
    \label{fig:IDD_lat_profiles_tank}
\end{figure}

First, three simple typical pencil beam algorithm benchmarking geometries were investigated. In all three cases, a tank (of dimensions of \numproduct{10 x 10 x 10} \unit{\cm^{3}}) was irradiated with a spot with nominal energy \SI{100}{\MeV}, an energy spread of \SI{1}{\MeV}, a spot size of \SI{0.3}{\cm}, an angular spread of \SI{1.0e-8}{\radian} and a correlation of \num{0}. The first case is the one in which the tank is composed homogeneously of water (\SI{0}{HU}). In the other two cases, a half-plane slab is introduced in the tank between 2 and 3 cm in depth in the upper-half of the x-y plane (with z being the depth). This is usually one of the most challenging geometries for pencil beam algorithms. In one case the slab was composed of bone-like tissue of \SI{1000}{HU} and in the other it was composed of air-like tissue of \SI{-1000}{HU}. The tank was created using an in-house DICOM CT scan writer and was composed of \numproduct{100 x 100 x 100} voxels with a voxel size of \numproduct{0.1 x 0.1 x 0.1} \unit{cm^{3}}. Two-dimensional slices of the dose distributions of YODA and TOPAS can be seen in Figure \ref{fig:2d_plots_tank}. Integrated depth doses (IDDs) and lateral profiles at different depths along the original spot axis can be seen in Figure \ref{fig:IDD_lat_profiles_tank}.

For these simple test cases, the visual agreement is excellent, as illustrated by both Figure \ref{fig:2d_plots_tank} and Figure \ref{fig:IDD_lat_profiles_tank}. This is also reflected in the 3D gamma index pass rates shown in Table \ref{tab:gamma_index_table} under the columns denoted by \SI{-1000}{HU}, \SI{0}{HU} and \SI{+1000}{HU}. The worst passing rate using the strict \SI{1}{\mm}, \SI{1}{\percent}, \SI{10}{\percent} dose cutoff is \SI{98.22}{\percent}. All passing rates presented can be further improved by fine tuning the splitting scheme. One way of doing so is to increase the number of rings. Another, is to take advantage of the underlying CT grid in the case of this perpendicular propagating spot. If in the lateral beam eye view grid one beamlet is placed per voxel and the spread is contained to the voxel lateral dimensions, the error is bound to decrease without much increase in computational cost.

\begin{table}[]
    \centering
    \begin{tabular}{|ccccccccc|}
        \hline
        \multicolumn{9}{|c|}{Gamma index}                                                                                                                                                                                                                                                 \\ \hline
        \multicolumn{3}{|c|}{Criteria} & \multicolumn{6}{c|}{Passing rates (\%) for}                                                                                                                                                                                                      \\ \hline
        \multicolumn{1}{|c|}{mm}       & \multicolumn{1}{c|}{\%}                     & \multicolumn{1}{c|}{\% - cutoff} & \multicolumn{1}{c|}{-1000 HU} & \multicolumn{1}{c|}{0 HU}  & \multicolumn{1}{c|}{+1000 HU} & \multicolumn{1}{c|}{H\&N}  & \multicolumn{1}{c|}{Prostate} & Lung  \\ \hline
        \multicolumn{1}{|c|}{1}        & \multicolumn{1}{c|}{1}                      & \multicolumn{1}{c|}{0}           & \multicolumn{1}{c|}{99.96}    & \multicolumn{1}{c|}{100}   & \multicolumn{1}{c|}{99.99}    & \multicolumn{1}{c|}{100}   & \multicolumn{1}{c|}{100}      & 100   \\
        \multicolumn{1}{|c|}{1}        & \multicolumn{1}{c|}{1}                      & \multicolumn{1}{c|}{10}          & \multicolumn{1}{c|}{98.22}    & \multicolumn{1}{c|}{99.93} & \multicolumn{1}{c|}{99.45}    & \multicolumn{1}{c|}{99.85} & \multicolumn{1}{c|}{99.58}    & 95.62 \\
        \multicolumn{1}{|c|}{2}        & \multicolumn{1}{c|}{2}                      & \multicolumn{1}{c|}{0}           & \multicolumn{1}{c|}{100}      & \multicolumn{1}{c|}{100}   & \multicolumn{1}{c|}{100}      & \multicolumn{1}{c|}{100}   & \multicolumn{1}{c|}{100}      & 100   \\
        \multicolumn{1}{|c|}{2}        & \multicolumn{1}{c|}{2}                      & \multicolumn{1}{c|}{10}          & \multicolumn{1}{c|}{99.61}    & \multicolumn{1}{c|}{99.95} & \multicolumn{1}{c|}{99.78}    & \multicolumn{1}{c|}{99.99} & \multicolumn{1}{c|}{99.99}    & 99.72 \\
        \multicolumn{1}{|c|}{3}        & \multicolumn{1}{c|}{3}                      & \multicolumn{1}{c|}{0}           & \multicolumn{1}{c|}{100}      & \multicolumn{1}{c|}{100}   & \multicolumn{1}{c|}{100}      & \multicolumn{1}{c|}{100}   & \multicolumn{1}{c|}{100}      & 100   \\
        \multicolumn{1}{|c|}{3}        & \multicolumn{1}{c|}{3}                      & \multicolumn{1}{c|}{10}          & \multicolumn{1}{c|}{99.73}    & \multicolumn{1}{c|}{100}   & \multicolumn{1}{c|}{99.85}    & \multicolumn{1}{c|}{99.99} & \multicolumn{1}{c|}{100}      & 99.86 \\ \hline
    \end{tabular}
    \caption{Gamma index passing rates for different criteria and test cases.}
    \label{tab:gamma_index_table}
\end{table}

\subsubsection*{CT based anatomies}

In addition to the simplified tank geometries, three real CT images were also tested. The H\&N scan was taken from the CORT dataset \citep{craftSupportingMaterialShared2014}, the prostate scan was taken from the cancer imaging archive \citep{yorkePelvicReferenceData2019} and the lung scan was taken from the Holland Proton Therapy Center \citep{pastorserranoArtificialIntelligenceRadiotherapy2023}. The used isocenter locations and gantry angles are not meant to be clinical and were chosen only due to their simplicity of set-up in TOPAS.

\begin{figure}[H]
    \centering
    \begin{subfigure}{\linewidth}
        \includegraphics[width=\linewidth, trim={0 0 0 0}, clip]{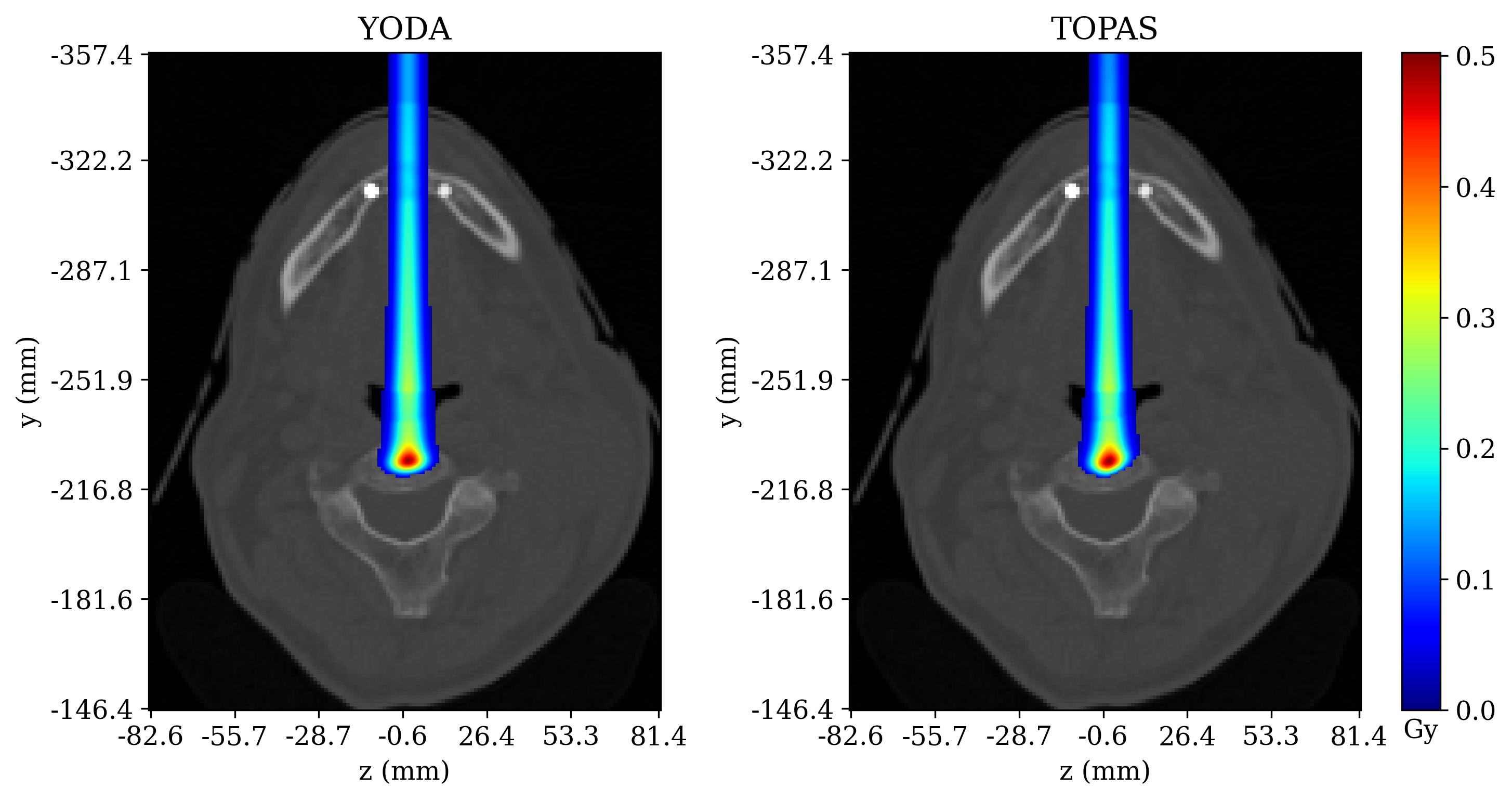}
        \caption{Head \& neck}
    \end{subfigure}
    \begin{subfigure}{\linewidth}
        \includegraphics[width=\linewidth, trim={0 0 0 0}, clip]{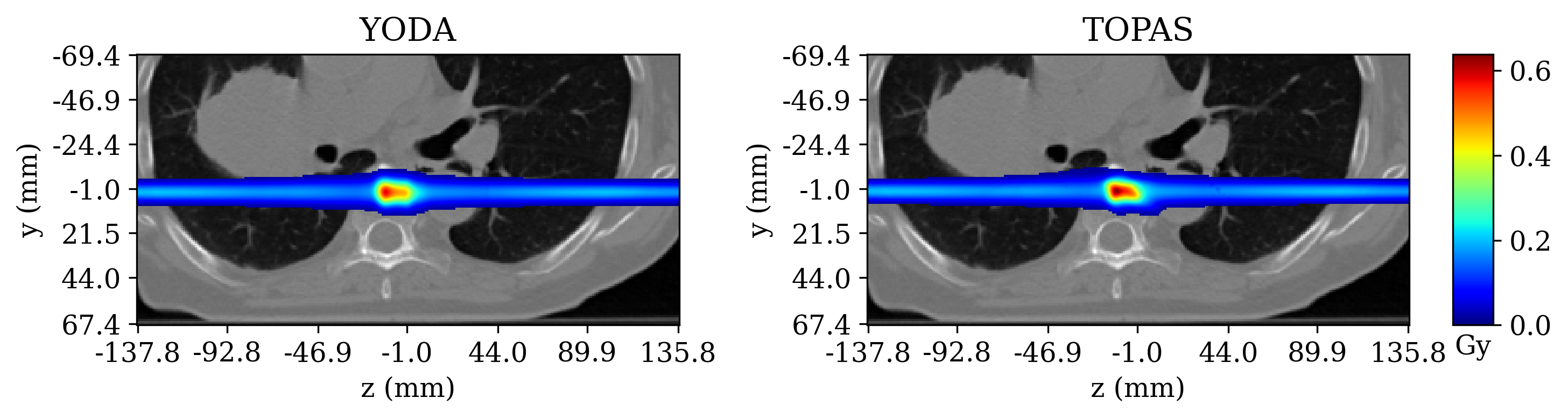}
        \caption{Lung}
    \end{subfigure}
    \begin{subfigure}{\linewidth}
        \includegraphics[width=\linewidth, trim={0 0 0 0}, clip]{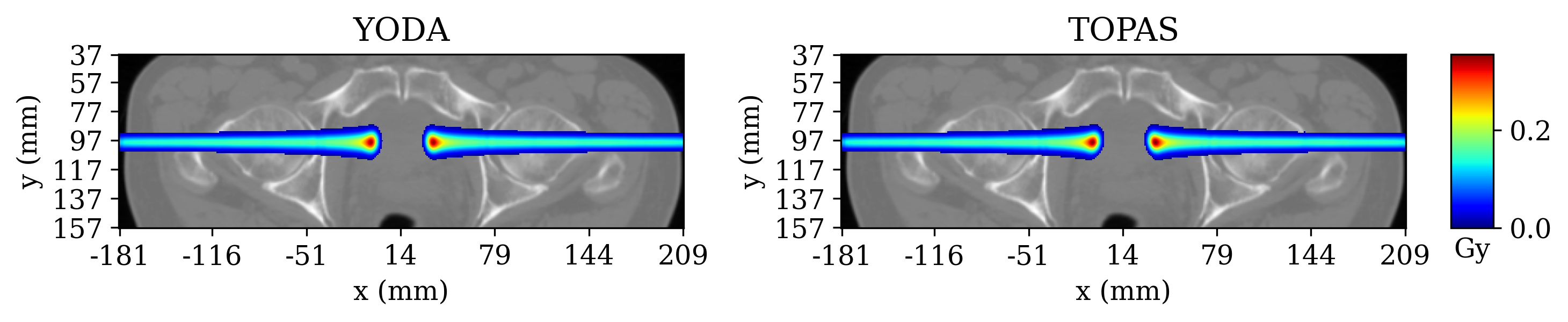}
        \caption{Prostate}
    \end{subfigure}
    \caption{2D plots for YODA and TOPAS in different CT images.}
    \label{fig:2d_profiles_CTs}
\end{figure}

\begin{figure}[H]
    \centering
    \begin{subfigure}{\linewidth}
        \centering
        \includegraphics[width=0.7\linewidth]{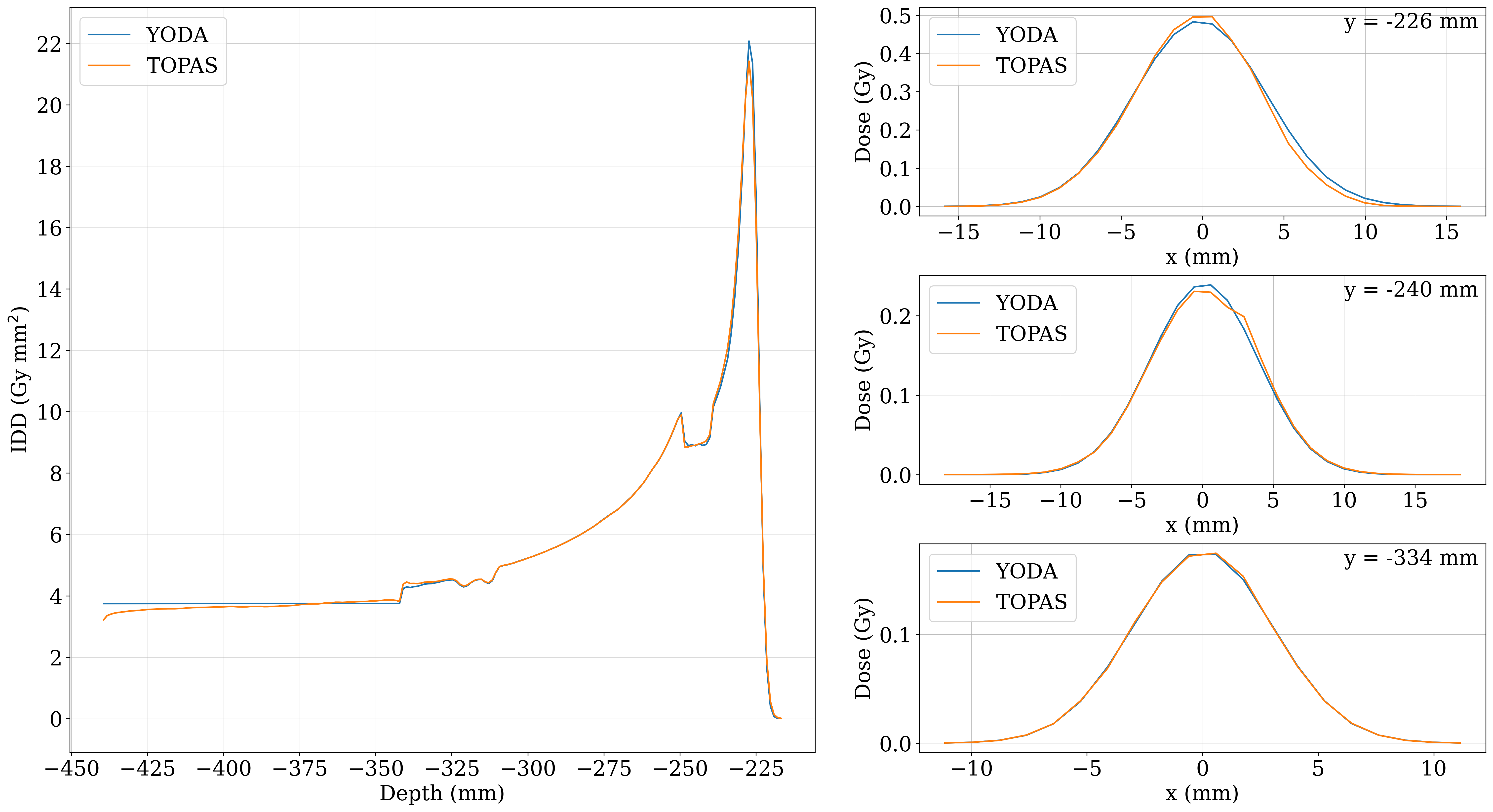}
        \caption{Head \& neck}
    \end{subfigure}
    \begin{subfigure}{\linewidth}
        \centering
        \includegraphics[width=0.7\linewidth]{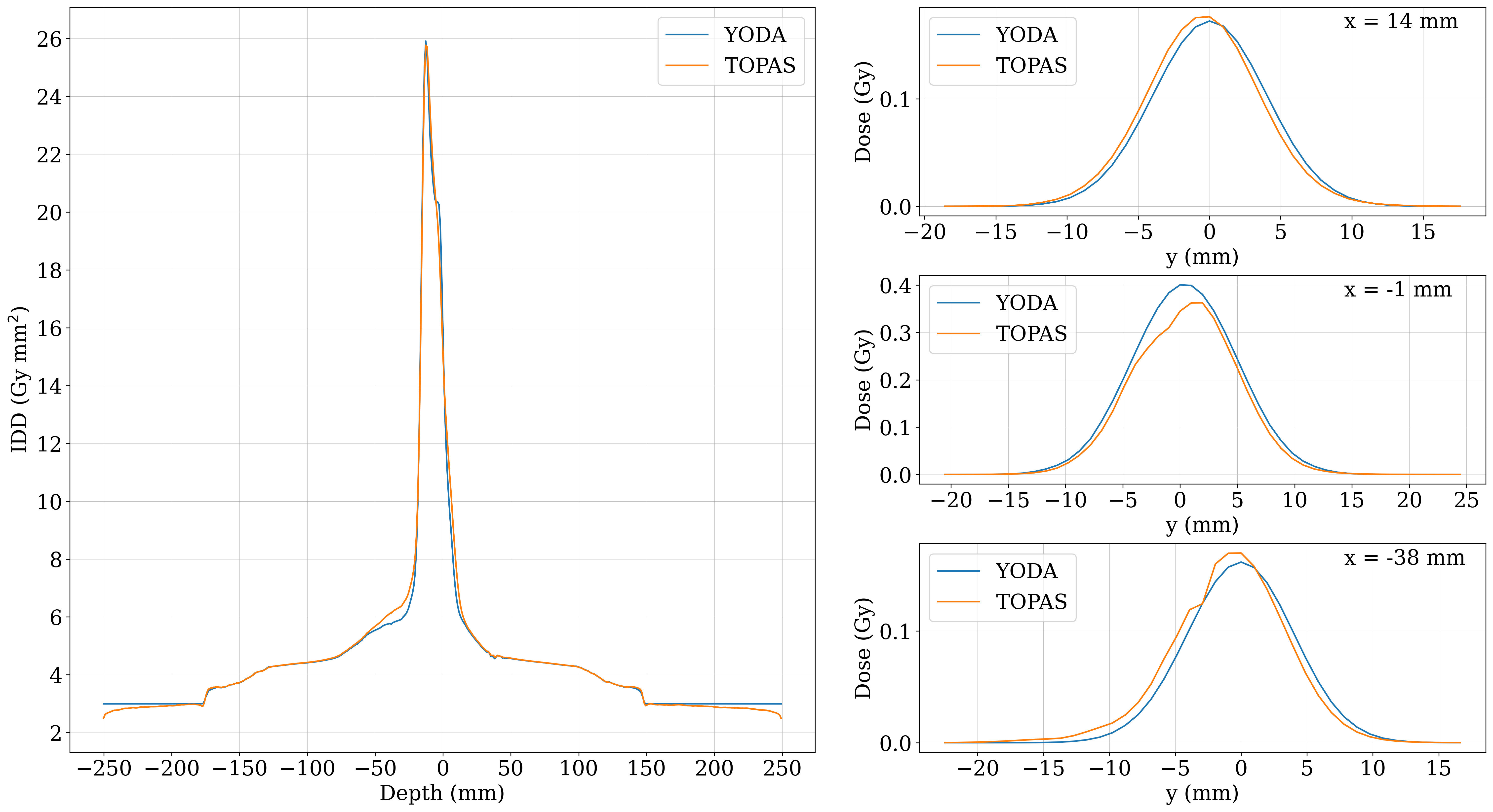}
        \caption{Lung}
    \end{subfigure}
    \begin{subfigure}{\linewidth}
        \centering
        \includegraphics[width=0.7\linewidth]{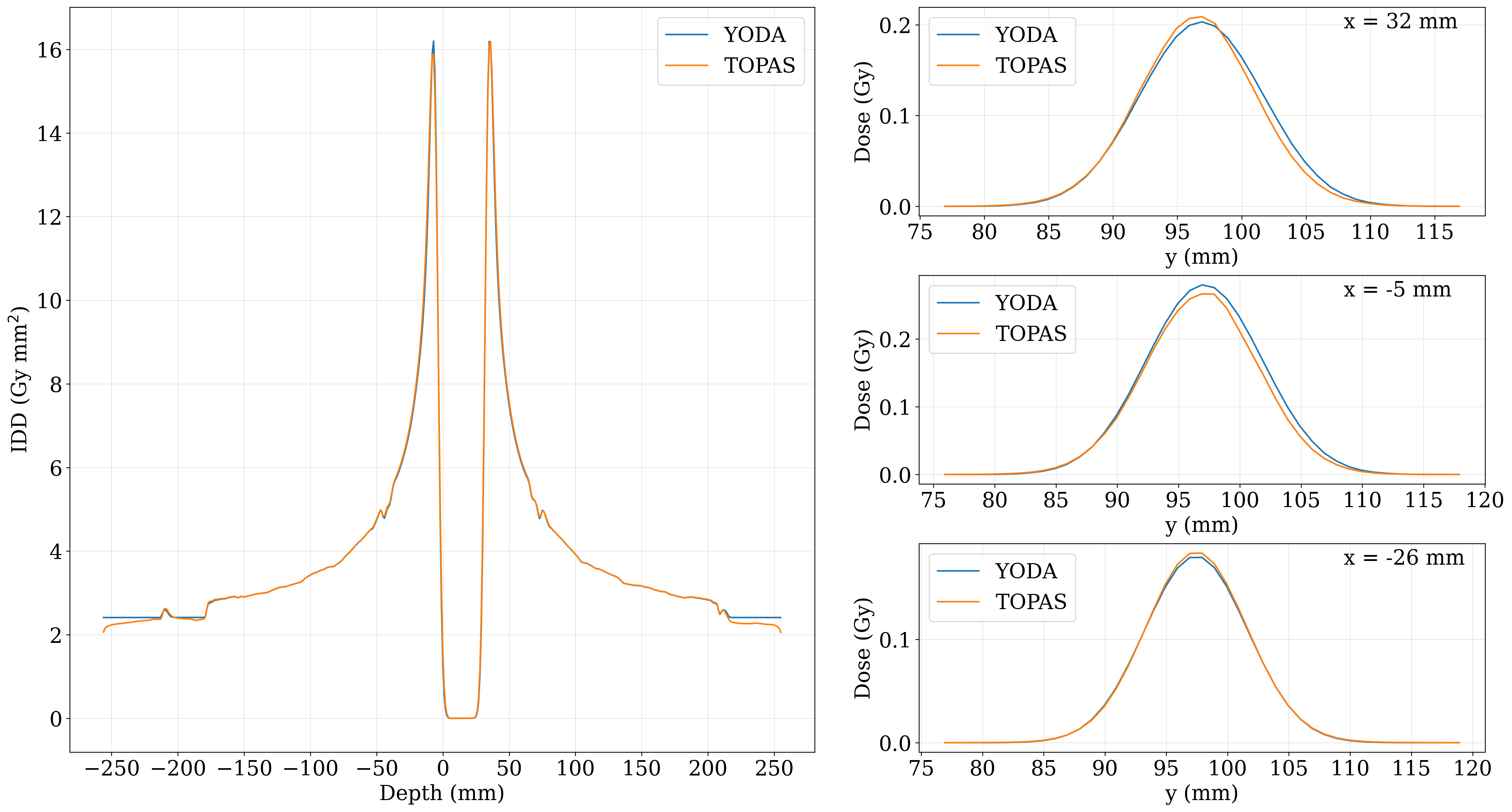}
        \caption{Prostate}
    \end{subfigure}
    \caption{IDD and lateral profiles for YODA and TOPAS in different anatomies.}
    \label{fig:IDDs_lat_profiles_CT_images}
\end{figure}

The H\&N scan, was irradiated with one spot that propagated from -y to +y (i.e., at a gantry angle of \SI{0}{\degree}) with a nominal beam energy of \SI{125}{\MeV} with the isocenter being the center of the CT scan volume. The two dimensional dose profile can be seen on the top row of Figure \ref{fig:2d_profiles_CTs} and the IDD and lateral profiles at three depths can be seen in the top row of Figure \ref{fig:IDDs_lat_profiles_CT_images}. Good agreement is observed, as the \SI{99.85}{\percent} gamma index pass rate from the H\&N column of Table \ref{tab:gamma_index_table} also shows. Figure \ref{fig:IDDs_lat_profiles_CT_images} shows a discrepancy in the air region between \SI{-440}{\mm} and \SI{-340}{\mm}. This is also the case for the lung and prostate cases visible in the middle and bottom rows of Figure \ref{fig:IDDs_lat_profiles_CT_images}. Two possible reasons are differences in the modelling of air between the two algorithms or a slight mismatch in the positioning of the beams with respect to the CT grid caused by the placement of the beam at the interface of voxels. Given that the agreement is good in the clinically relevant region of the scan this discrepancy is deemed acceptable.

The lung scan was irradiated with two spots where one goes from -x to +x and the other in the opposite direction (i.e., at \SI{90}{\degree} and \SI{270}{\degree} gantry angles respectively). Both spots had a mean energy of \SI{160}{\MeV}, energy spread of \SI{1}{\MeV}, a spot size of \SI{0.3}{\cm}, an angular spread of \SI{1.0e-8}{\radian} and a correlation of \num{0}. Given the challenging anatomy, the results from Figures \ref{fig:2d_profiles_CTs} and \ref{fig:IDDs_lat_profiles_CT_images} together with the passing rate of \SI{95.62}{\percent} from the lung column of Table \ref{tab:gamma_index_table} are very good. The lateral profiles from Figure \ref{fig:IDDs_lat_profiles_CT_images} show a consistent lateral shift between YODA and TOPAS at the \SI{-30.3}{\mm} and \SI{-79.1}{\mm} depths. A reason for this could be the initial location of the Gaussian split sub-spots on the CT scan surface. The spots are generally not aligned with the CT grid (as such alignment is only possible in cases of perfectly perpendicular beams) and therefore slight asymmetries could arise if spots are placed exactly at the interface of voxels. The accuracy can be improved by fine-tuning the Gaussian beam splitting scheme in several ways. One is to include the number of rings and the number of beamlets per ring into the optimization procedure itself. Another is to consider alternative, non-concentric sub-spot arrangements. A metric for lateral heterogeneity could help in guiding the optimization towards sparsely placing beamlets in areas of low heterogeneity and more densely covering areas with high heterogeneity. Lastly, a progressive splitting scheme could also be employed, whereby once a threshold of lateral heterogeneity has been reached, the beamlets encountering it are re-consolidated and a new (finer) split occurs. Given that the parameters of such schemes can be pre-optimized and tabulated the computational increase of such an approach could be kept minimal.

The prostate case set-up was identical to that of the lung with the only difference being the spot mean energy of \SI{165}{\MeV} and the spot energy spread of \SI{0.825}{\MeV}. Here again the agreement is very good as seen in Figures \ref{fig:2d_profiles_CTs} and \ref{fig:IDDs_lat_profiles_CT_images} and by the high passing rate of \SI{99.58}{\percent} from the prostate column of Table \ref{tab:gamma_index_table}.

\subsection{Dose change computations}

In addition to the dose engine performance, the performance of the dose change computation was also benchmarked. Given a specific volume within the CT scan denoted as ROI, the adjoint component is able to cheaply and accurately compute the change in the dose deposited in the ROI (for small enough anatomical perturbations). The speed of such an operation far exceeds that of plain re-computation as effectively, the only computation necessary comes in the form of vector inner products. This could be employed in an online re-adaptation trigger system where YODA assesses the effect of delivering yesterday's plan on today's anatomy. The benchmark starts with the same simplified tank test-cases and thereafter moves toward more realistic cases using RT plans for clinical RT structures on CT images.

\subsubsection*{Simplified tank geometries}

\begin{figure}[H]
    \centering
    \includegraphics[width=0.55\linewidth]{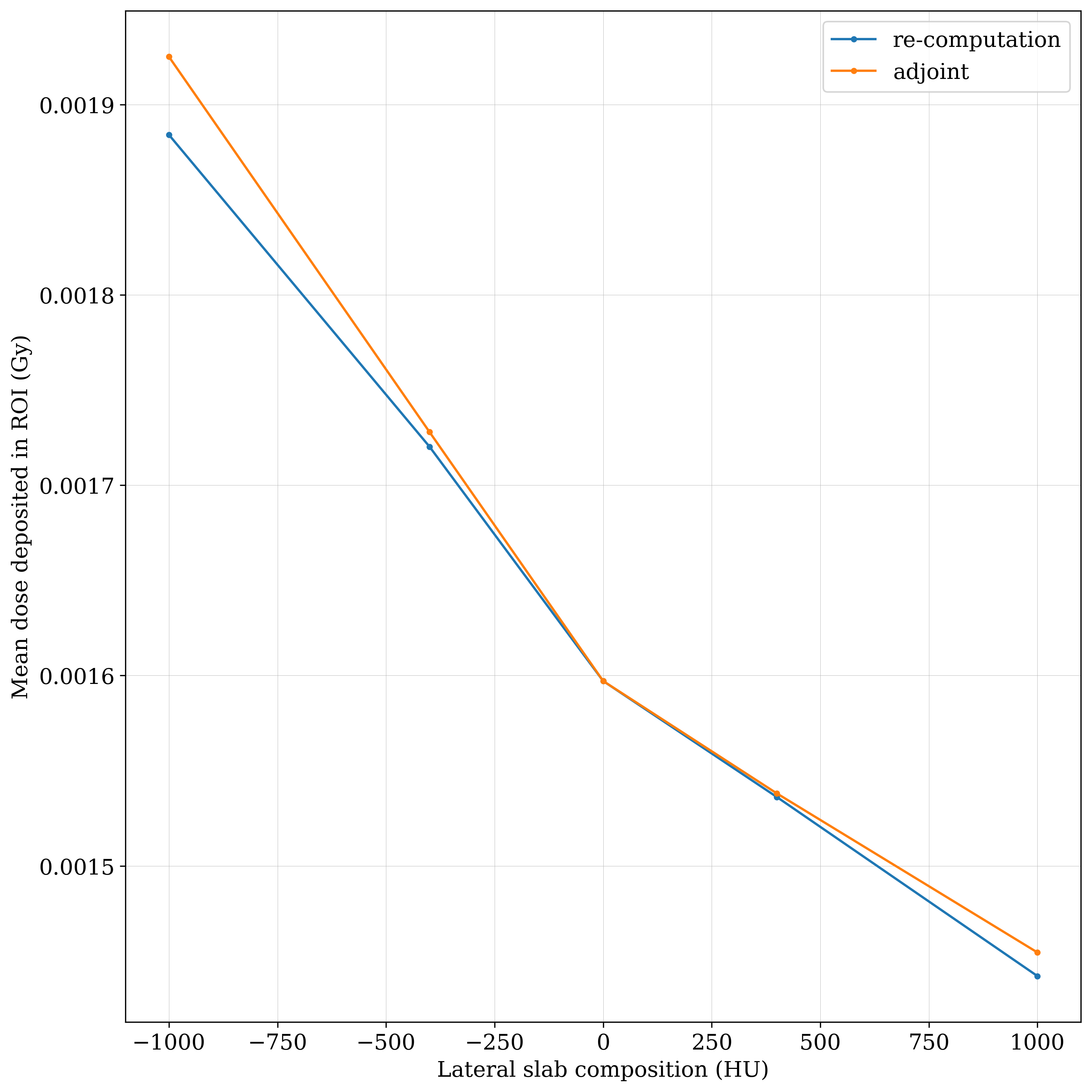}
    \caption{Re-computed versus adjoint computed doses for the simplified tank geometries.}
    \label{fig:adjoint_vs_recomputation_tank}
\end{figure}

In the case of the simple tank geometries, the adjoint component used a ROI defined as everything past the depth of \SI{60}{\mm} in the tank. The composition of the half-slab was varied from \SI{-1000}{HU} to \SI{+1000}{HU}. The mean dose deposited in the ROI was computed for each new geometry using two methods: re-computations and adjoint computations. Figure \ref{fig:adjoint_vs_recomputation_tank} shows the mean dose deposited in the ROI as a function of the HU composition of the slab. The two lines are close one to another around the value of 0 HU which was considered the base case and they start to diverge towards the edges of the HU domain. The maximal relative error of \SI{2.2}{\percent} occurs at the \SI{-1000}{HU} end of the HU domain. Based on these results, it can be concluded that the adjoint component is capable of cheaply and accurately computing the change in the deposited dose in the ROI for this test case.

\subsubsection*{Treatment plan tests}

Two treatment plans for the GTV were generated for two H\&N patients (patients 1 and 2) in Raystation \citep{bodensteinerRayStationExternalBeam2018}. The plans are not clinical and are only used for the purpose of creating conformal doses around the target. Patient 1 had a non-robust plan with 1031 spots and patient 2 had a robust plan with 344 spots. Both plans were split according to a $1+6+6+12$ Gaussian beam splitting scheme. Each patient had multiple repeat CTs (rCTs) which were registered to the pCT using the simple-itk library \citep{beareImageSegmentationRegistration2018}. The adjoint component computed the change in the GTV dose caused by the new CT image. This is meant to simulate the situation of a daily re-adaptation trigger system where the effect of yesterday's plan is assessed on today's anatomy. As long as the anatomical changes between the planning and repeat CT images are not too large, the adjoint component is accurate and fast as it does not require re-computing the original plan on the new image.

\begin{figure}[H]
    \centering
    \includegraphics[width=.85\linewidth]{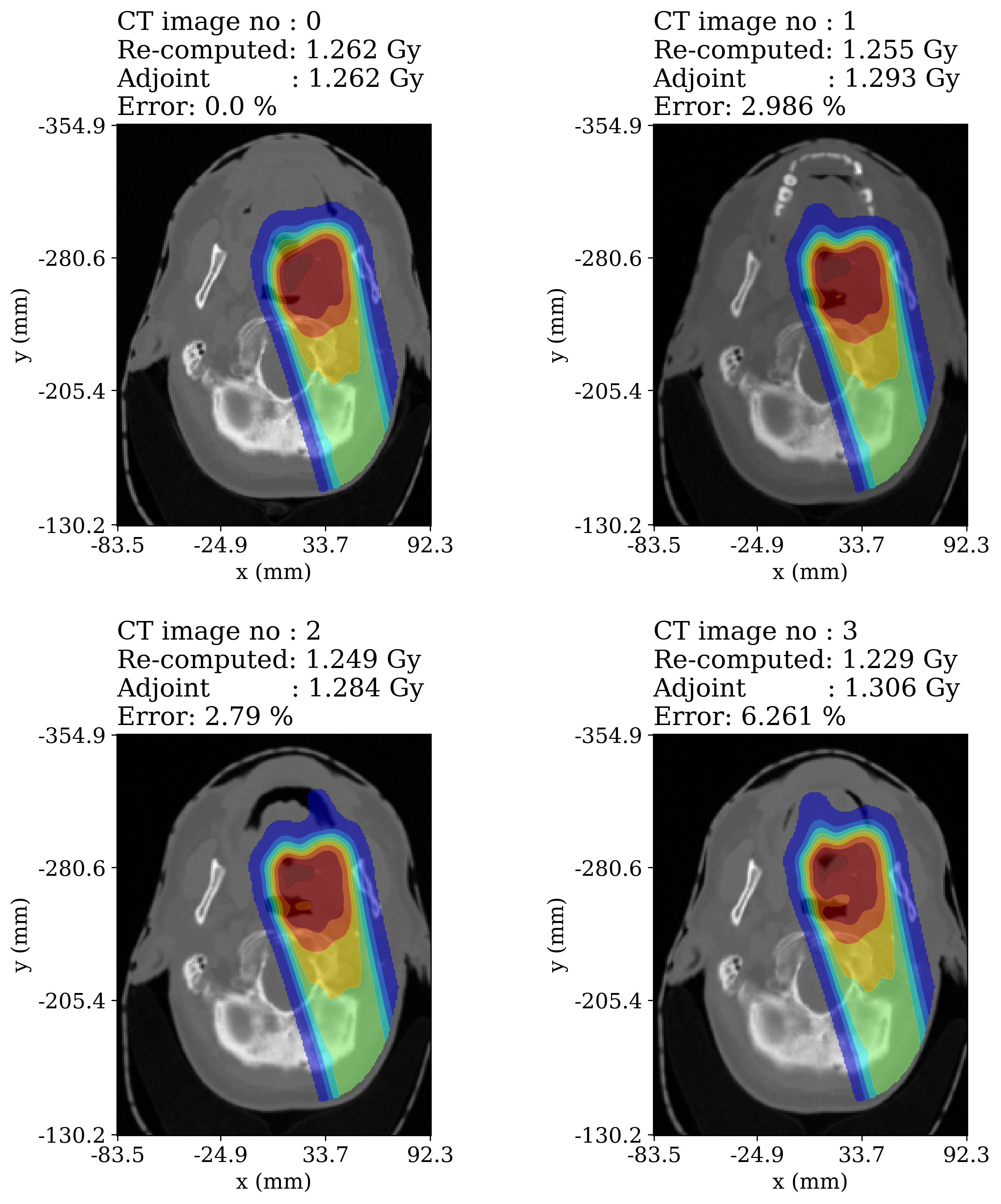}
    \caption{Re-computed doses on rCTs and adjoint prediction for dose to the GTV for Patient 1.}
    \label{fig:pat_19_adjoint}
\end{figure}

Figure \ref{fig:pat_19_adjoint} shows each of the CT images for patient 1 (image number 0 is the planning image), a 2D profile of the re-computed dose distribution on the CT image, the GTV dose computed via re-computation and via the adjoint component and the relative error between these two results. In this case, the adjoint component attains a maximal error of \SI{6.2}{\percent}. Thus, despite the plan not being robustly optimized, the adjoint component is capable of avoiding an expensive re-computation attaining an acceptable error.

\begin{figure}[H]
    \centering
    \includegraphics[width=.85\linewidth]{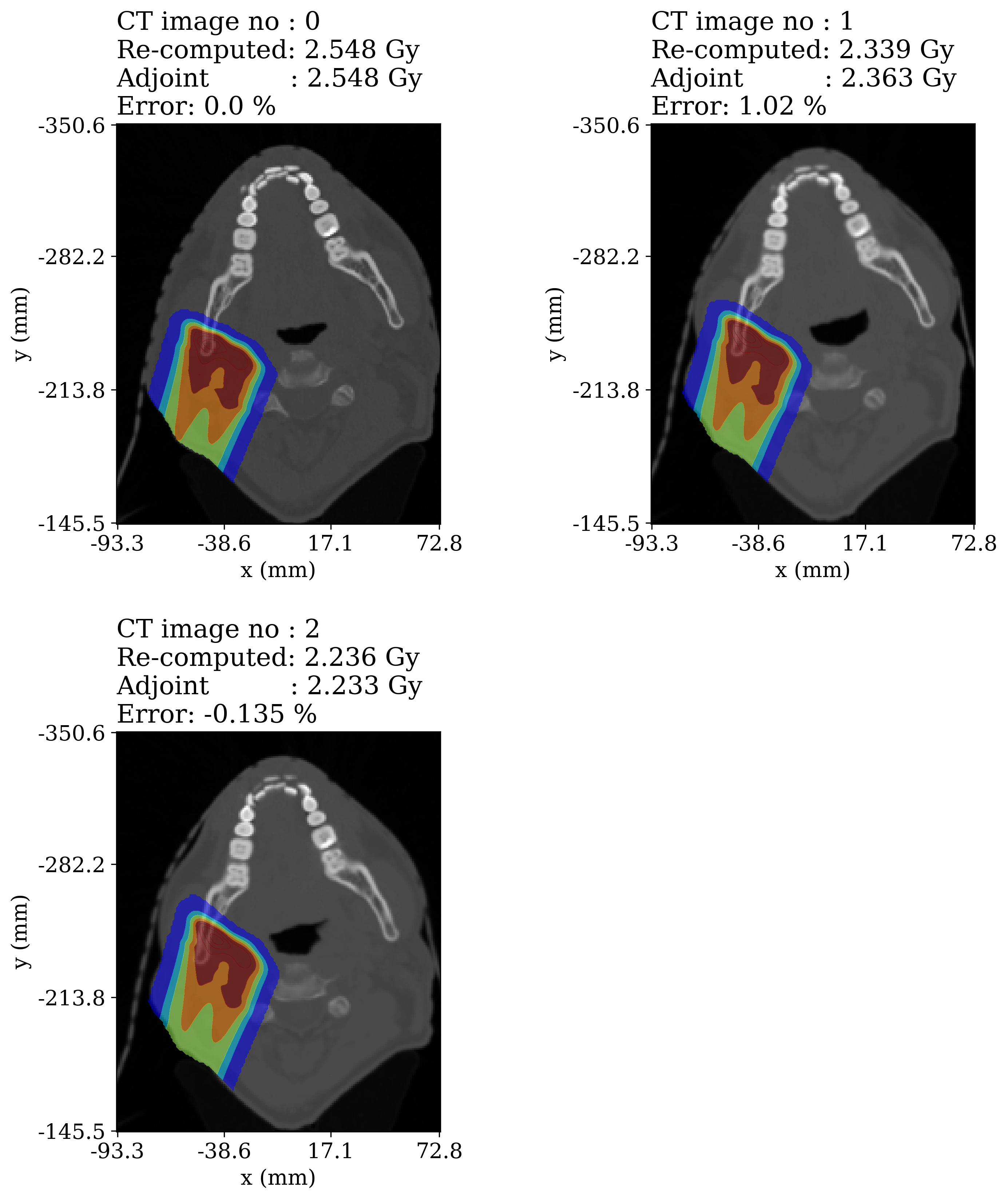}
    \caption{Re-computed doses on rCTs and adjoint prediction for dose to the GTV for patient 2.}
    \label{fig:pat_163_adjoint}
\end{figure}

Figure \ref{fig:pat_163_adjoint} shows for patient 2 the re-computed doses overlaid on the corresponding CT images, the re-computed and adjoint computed doses to the GTV and the relative error between these two results. In the case of this robustly optimized plan, the maximal error is \SI{1}{\percent}.

\section{Conclusions}
\label{sec:conclusion}

In this work Yet anOther Dose Algorithm (YODA) and its performance in a variety of test cases was presented. YODA uses a hybrid approach to solve a physics-based approximation to the same equations that MC methods solve. This approach enables YODA to achieve TOPAS like performance with a significant speed-up. The lowest three dimensional gamma index passing rates achieved using the strict criteria of \SI{1}{\mm}, \SI{1}{\percent}, \SI{10}{\percent} cut-off is \SI{95.62}{\percent} in the lung case. YODA computes a treatment plan spot in \SI{2}{\second} while the same spot takes hours in TOPAS. If the speed would be further improved (e.g., via a GPU implementation), YODA could be used as a patient-specific quality assurance tool by tapping into the data stream between the TPS and the delivery machine to quickly re-construct the dose to be delivered. Alternatively, the logfiles could be used after treatment to re-construct the actually delivered dose to the patient. A multi-treatment site patient cohort study is necessary to validate the accuracy of YODA versus commercial TPS calculations in a wide variety of settings. Additionally, nuclear interactions must be accounted for. However, given that the dose engine contained in Eclipse (AcurosPT) is accurate with criteria of \SI{2}{\mm}, \SI{2}{\percent} in hetereogeneous cases \citep{demartinoDoseCalculationAlgorithms2021} and the various speed and accuracy improvements still achievable in YODA it can be concluded that this engine could compete with/replace other commercial dose algorithms and is certainly capable of TPS independent dose calculations.

Next to performing TPS independent dose calculations, YODA can leverage the adjoint component to accurately compute dose changes caused by small enough anatomical changes. Such a feature, to the best of the authors' knowledge, has not been integrated into a dose algorithm before. This component could be used in a time constrained re-adaptation trigger system where on the given day YODA avoids re-computing the old treatment plan on the new CT image if the CT image is deemed anatomically close enough to the original one. This performance was illustrated via two treatment plans where a maximal error of \SI{6.2}{\percent} was achieved for a non-robustly optimized plan and \SI{1}{\percent} for a robustly optimized plan. Alternatively, if log-files would be available during treatment delivery YODA would be capable of halting erroneous deliveries in near real-time (i.e., below energy layer switching times) by converting spot position differences into anatomical changes and ultimately into dosimetric changes via the adjoint component.

\section{Conflicts of interest and acknowledgements}

The authors wish to acknowledge that the manuscript is partly funded by Varian, a Siemens Healthineers Company.
The authors declare that they have no known competing financial interests or personal relationships that could have appeared to influence the work reported in this paper. Moreover, no data was used for the research described in the article.

\section{Credit statement}

\textbf{Tiberiu Burlacu:} Conceptualization, methodology, software, validation, formal analysis, data curation, investigation, writing - original draft, writing - review \& editing, visualization. \\
\textbf{Danny Lathouwers:} Conceptualization, methodology, software, validation, resources, writing - review \& editing, supervision. \\
\textbf{Zolt\'{a}n Perk\'{o}:} Conceptualization, methodology, validation, resources, writing - review \& editing, supervision, project administration, funding acquisition.

\bibliographystyle{agsm}
\bibliography{./bib/bibliography.bib}

\end{document}